\documentclass[10pt]{article}

\usepackage{amsmath}
\usepackage{amssymb}

\usepackage{graphicx}

\usepackage{cite}
\usepackage{rotating}

\usepackage{color} 

\usepackage[labelfont=bf,labelsep=period,justification=raggedright]{caption}

\makeatletter
\renewcommand{\@biblabel}[1]{\quad#1.}
\makeatother

\date{}

\begin{document}

\begin{flushleft}
{\Large
\textbf{Inductive Game Theory and the Dynamics of Animal Conflict}
}
\\
Simon DeDeo$^{1,2}$, 
David C. Krakauer$^{1}$, 
Jessica C. Flack$^{1,3,\ast}$
\\
\bf{1} Santa Fe Institute, Santa Fe, NM 87501, USA 
\\
\bf{2} Institute for the Physics and Mathematics of the Universe, University of Tokyo, Kashiwa-shi, Chiba 277-8582, Japan 
\\
\bf{3} Yerkes National Primate Research Center, Emory University, Atlanta, GA 30322, USA\\
$\ast$ E-mail: jflack@santafe.edu
\end{flushleft}

\section*{Abstract}
Conflict destabilizes social interactions and impedes cooperation at multiple scales of biological organization. Of fundamental interest are the causes of turbulent periods of conflict. We analyze conflict dynamics in a monkey society model system. We develop a technique, Inductive Game Theory, to extract directly from time-series data the decision-making strategies used by individuals and groups. This technique uses Monte Carlo simulation to test alternative causal models of conflict dynamics. We find individuals base their decision to fight on memory of social factors, not on short timescale ecological resource competition. Furthermore, the social assessments on which these decisions are based are triadic (self in relation to another pair of individuals), not pairwise. We show that this triadic decision making causes long conflict cascades and that there is a high population cost of the large fights associated with these cascades. These results suggest that individual agency has been over-emphasized in the social evolution of complex aggregates, and that pair-wise formalisms are inadequate. An appreciation of the empirical foundations of the collective dynamics of conflict is a crucial step towards its effective management. 

\section*{Author Summary}
Persistent conflict is one of the most important contemporary challenges to the integrity of society and to individual quality of life. Yet surprisingly little is understood about conflict. Is resource scarcity and competition the major cause of conflict, or are other factors, such as memory for past conflicts, the drivers of turbulent periods? How do individual behaviors and decision-making rules promote conflict? To date, most studies of conflict use simple, elegant models based on game theory to investigate when it pays to fight. Although these models are powerful, they have limitations: they require that both the strategies used by individuals and the costs and benefits, or payoffs, of these strategies are known, and they are tied only weakly to real-world data. Here we develop a new method, Inductive Game Theory, and apply it to a time series gathered from detailed observation of an actual primate system. We are able to determine which types of behavior are most likely to generate periods of intense conflict, and we find that fights are not explained by single, aggressive individuals, but by complex interactions among groups of three or higher. Understanding how memory and strategy affect conflict dynamics is a crucial step towards designing better methods for prediction, management and control. 

\section*{Introduction}
 
Conflict is dissipative. Organisms, aggregates, and societies must overcome the destabilizing consequences of conflict in order to persist \cite{Buss1987, Michod2000, Burt2006, Flack2005b, Flack2006}. Conflict consequently plays a central role in the evolution of social organization. Particularly problematic for social stability is ontogenetic conflict  -- conflict that finds expression in fights between individuals over the course of their lifetimes. Much is understood about control mechanisms \cite{Dreber2008, Clutton1995, Frank2003, Flack2006, Flack2005, Flack2005b, Waal2000, Aureli2000}, factors driving escalation of pair-wise contests \cite{Clutton1979, Maynard1974, Parker1981, Taylor2003}, the influence of third parties on conflict outcome through coalition formation \cite{Mesterton2006, Noe1995}, audience \cite{Johnstone2001, Covas2007} and reputation effects \cite{Nowak1998}, and redirected aggression \cite{Karzem2005}. Somewhat paradoxically, less is understood about the causes of conflict, and almost nothing is known about the dynamics of multiparty conflict --conflicts that spread to involve more than two individuals to encompass a sizable fraction of a group. Multiparty conflicts are common in many gregarious individual societies \cite{Harcourt1992}. In these systems, it is often difficult to establish why individuals become involved in an ongoing fight or why the fight started.

 A standard assumption is that individual strategies are highly tuned to resource competition. Under this assumption the cause of any single conflict is immediate competition for resources over short time intervals. The probability of fighting depends directly on the payoff obtained from acquiring the resource in the present. These resources can include food, mates and dominance status. The latter is thought to improve access to food and mates.  However, individual memory for previous interactions can alter the occurrence and course of future conflicts, promoting longer-timescale, competitive dynamics. This is because memory for regular patterns of past conflict facilitates prediction of future conflict, allowing individuals to respond strategically. It is well understood, for example, that competition for dominance between a pair of individuals can be played out over many months and involve alliances and coalitions \cite{Waal1982}.  Memory can also introduce costs, as it can lead to the amplification of conflict or to the eruption of a sequence of related fights: a ``cascade" \cite{Boehm1987, Cairns2003}. Such turbulent periods can increase the probability of injury and stress, both of which are associated with increased mortality \cite{sapolsky2005}. Large conflicts can increase the probability that  individuals not involved in an initial dispute will be drawn in, and so can increase the ``population cost" of conflict. Thus critical questions include: how do individuals decide to fight, are multiparty conflicts are reducible to pair-wise interactions or do they involve irreducible higher-order interactions. How do alternative decision-making rules, or strategies, effect inter-conflict dynamics and organizational stability, and what role does memory play in amplifying and dampening conflict?
 
Addressing these questions in multiplayer systems requires models that make few or no assumptions about payoffs, as these are rarely known, and which are tractable when allowing for higher-order interactions (more than pairwise interactions). In standard game theory models -- a canonical approach to the study of conflict -- payoffs are posited, higher order strategic interactions are typically neglected, and data rarely derive from temporally resolved, natural observations of strategic interactions. The goal is to provide solution concepts for games that find uninvadible strategies, rather than to extract from the data directly those strategies individuals are playing \cite{Hammerstein2006}. 

To complement these standard \emph{deductive} game theoretic approaches, we introduce {\it Inductive Game Theory}, in which the strategies used by individuals, and their consequences for social dynamics, are derived computationally from highly resolved time-series data on competitive processes.  Methodologically, this approach borrows from statistical inference methods now standard in genetics. The goal in genetics is typically to reconstruct gene interactions from expression-profile, time-series. The problem is that the number of transcripts is usually far greater than the number of independent observations. Hence priors need to be imposed on permissible solutions. The goal of Inductive Game Theory is to extract decision-making strategies and behavioral time series from known interaction networks. Hence these problems are in some sense inverse of each other.  In the Discussion, we return to this issue, expanding the scope of IGT to consider non-conflict time series.

In the body of the paper we develop the inductive game theory approach and apply it to a conflict data set from a pigtailed macaque ({\it Macaca nemestrina}) group. The macaque ({\it Macaca}) genus and its subset species are natural model systems for studying the role of complex conflict dynamics in social evolution. This is because in macaque societies individual decision making is plastic and guided by learning, conflict is frequent and typically involves multiple, unrelated players, and social dynamics occur over multiple timescales \cite{Flack2004, Thierry2004}. The particular pigtailed macaque group we study contains 48 socially-mature individuals (84 individuals in total) housed socially in a large compound at the Yerkes National Primate Research Center Field Station in Lawrenceville, Georgia (see Empirical Methods.) Data were collected over a series of four months in which the group was stable (no reversals in dominance status). Conflict events -- ``fights'' -- in this group vary in duration, number of participants, and other measures of severity \cite{Flack2005, Flack2005b}.  Because the entire sequence of conflict events was collected, including data on fight duration, participant identity, and participant behavior, we are able to construct a highly-resolved time-series for each observation period. A total of 1,096 fights in 158 hours were observed over the study period; the names of individuals in each fight were recorded. 

\section*{Time Series Correlations}
\label{correlations}

\begin{figure}[!ht]
\begin{center} \includegraphics[width=4in]{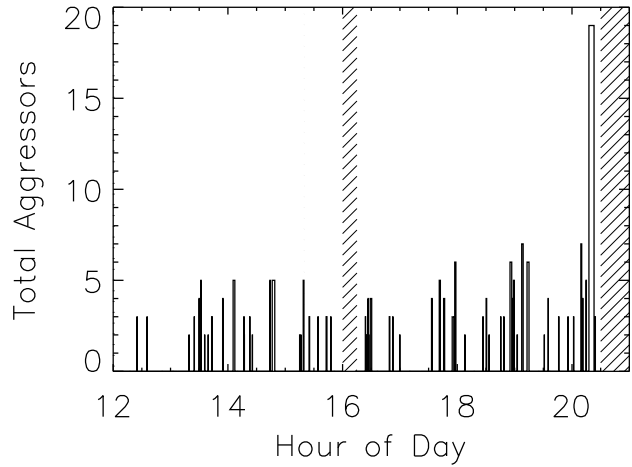} \end{center}
\caption{{\bf Conflict event time-series data from one observation period.} Begins at 12:00 hours, and ends just after 20:00 hours. Plotted on the y-axis as ``Total Fight Size'' is the number of conflict participants per conflict, regardless of whether the participant was an aggressor, recipient, or intervener. The graph gives a sense of the distribution of conflict sizes, and conflict lengths, and the distribution of intervening peaceful periods. Hatched bars indicate periods without data collection.}
\label{day_five}
\end{figure}

We begin by asking whether fight sizes are correlated in time. An example time-series, from a single eight-hour observation period and showing fight size and duration, is in Fig.~\ref{day_five}; one may construct from this various autocorrelation functions. Surprisingly, the sizes of fights are nearly uncorrelated over the course of the day. Larger-than-average fights do not, for example, predict the appearance of larger-than-average fights later. This is discussed in greater detail in the Supporting Information.

We then ask whether there are correlations across fights in membership -- does the the appearance of individual $A$ in a fight at time step one predict the appearance of $B$ in the next fight? For simplicity, the only within-fight information we use is individual identity data; we do not take into account individual behavior (\emph{e.g.}, aggressor, recipient, intervener, and so forth -- see Empirical Methods), nor do we consider which individuals interacted within fights. Given this, the simplest correlations we can observe are correlations in membership across fights separated by one peace bout.  We write these $P(A\rightarrow B)$, estimated as $N(B|A)/N(A)$: the number of fights involving $B$ that followed a fight involving $A$, divided by the number of fights involving $A$. Informally, $P(A\rightarrow B)$ gives the probability of observing $B$ in a conflict given that one has just observed a conflict involving $A$. 

\begin{figure}[!ht] 
\begin{center} \includegraphics[width=4in]{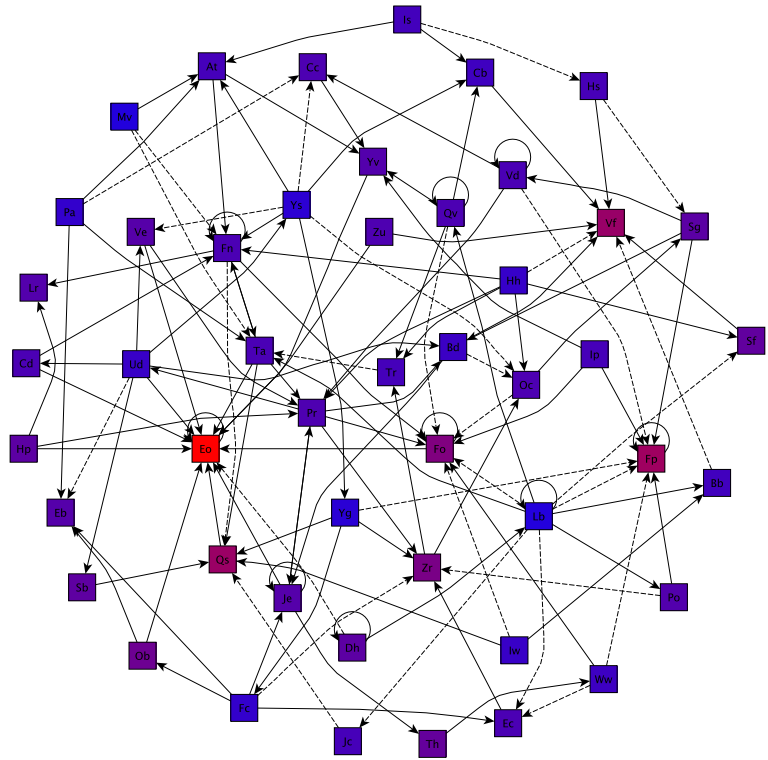} \end{center} 
\caption{{\bf The network of the strongest correlations detected in the data set, shown as directed edges between individuals.} $\Delta P$ (as defined in Eq. \ref{defdelta}) positive is denoted as a solid line, and $\Delta P$ negative as a dashed line; arrows denote the forward direction of time. Edges with $|\Delta P|$ above 6\% (at 95\% confidence) are shown; note that detections of different edges are not independent. Node color indicates frequency, with blue meaning rare in fights, and red, frequent.}
\label{pabgraph}
\end{figure}

The probabilities will vary for different pairs of individuals. In order to remove time-independent effects on individual participation in fights, we compute $\Delta P(A\rightarrow B)$; the difference between the null-expected $P$ and that measured from the data:
\begin{equation}
\label{defdelta}
\Delta P(A\rightarrow B)=\frac{N(B|A)-N_{\mathrm{null}}(B|A)}{N(A)},
\end{equation}
where $N_{\mathrm{null}}(B|A)$ is the average from a large Monte Carlo set of null models generated by time-shuffling the series but not shuffling identities within fights. Fig.~\ref{pabgraph} shows some of the strongest correlations of this form found between the 48 individuals, in the form of a directed graph.

Correlations across fights can also be generated by subgroups deciding to fight in response to other subgroups fighting previously. There are several possible variations in subgroup-generated correlations. We consider only the two computationally simplest correlational structures. Correlations of the form $\Delta P(AB\rightarrow C)$ reveal the extent to which the presence or absence of a pair of individuals at one time predicts the appearance of a particular individual at the next step. They can be defined:
\begin{equation}
\label{pabcdef}
\Delta P(AB\rightarrow C)=\frac{N(C|AB)-N_{\mathrm{null}}(C|AB)}{N(AB)},
\end{equation}
as can $\Delta P(A\rightarrow BC)$, the extent to which the presence of an individual at the previous step predicts the presence of a pair at the next step:
\begin{equation}
\label{pabc_2_def}
\Delta P(A\rightarrow BC)=\frac{N(BC|A)-N_{\mathrm{null}}(BC|A)}{N(A)},
\end{equation}
Using this combinatorial Monte Carlo technique, we find significant correlations for both these structures. Plots of the distribution of these three correlations can be found in the Supporting Information. As one measures higher-level correlations, between, for example, triplets and individuals, the effective sample size -- number of relevant observations (the conditional $N(x|y)$ and $N_\mathrm{null}(x|y)$) -- drops, while the number of parameters to estimate rises combinatorically. This leads to a rapid decrease in signal-to-noise on any one observable.

Extracting overall significance levels for $\Delta P$ measurements requires caution. For example, since individuals are correlated within fights, and these correlations are maintained by the null model, the various $\Delta P$ measurements are not independent of each other. A Monte Carlo simulation of the expected numbers of correlations confirms that the excess of positive $\Delta P$ values in the observed data is significant at $p<0.001$; these issues are discussed in greater detail in the Supporting Information. 

A similar analysis can be done for two-step correlations between named individuals and groups; while individual detections can be made, Monte Carlo simulations of the expected noise properties of the $\Delta P$ measurement suggest that such correlations, should they exist, are too weak to detect even in the full sample of 1096 fights.

\section*{The Space of Strategies}
\label{stratdef}

Given the observed correlations, we now consider the causal mechanisms underlying the detectable individual and subgroup correlations. To do so, we introduce a class of minimal models for social reasoning, or ``strategy space.'' Full specification of these models is in the Supporting Information section, ``Simulation Specification.''

We suppose that each individual or subgroup decides whether to join a fight based on composition of the previous fight. The space of possible strategies can then be written as $\mathcal{C}(n,m)$, where $n$ is the size of the relevant group in the previous fight, and $m$ is the number of individuals making the decision. We allow decisions to be probabilistic (``mixed,'' in the game theory terminology), so that a particular fight composition can lead, with some probability distribution, to different kinds of subsequent fights.

Each element of a $\mathcal{C}(n,m)$ strategy is a number between $+1$ and $-1$, specifying the probability that the appearance of a particular $n$-tuple leads to a recommendation that a particular $m$-tuple join, or avoid, the next fight. These probabilities derive directly from the data, using the equations given in the previous section to determine whether there is a significant identify correlation across fights between two individuals or pairs. A negative value can be interpreted as repulsion or inhibition, and a positive value can be interpreted as attraction or stimulation.

In the case that a particular $m$-tuple receive multiple, possibly incompatible, recommendations to join or avoid, which is always possible because each fight has a minimum of two individuals involved, the decision to join or avoid can be resolved by introducing a \emph{temperament} parameter, which we call a ``combinator.'' We choose {\tt AND} and {\tt OR} to capture the two ends of the spectrum of individual temperaments. Under the conflict averse, or conservative {\tt AND} combinator, an $m$-tuple must receive recommendations to join from all relevant $n$-tuples. Under the maximally conflict-prone {\tt OR} combinator, a single recommendation to join is sufficient.

We begin with a randomly generated, spontaneous ``seed'' pair. These seeds can trigger a subsequent series of fights (a ``cascade'')  that in our simulation build up a time series. At some point, a particular fight may lead to no recommendations to join, or a recommendation that only a single individual join; at this point, the cascade ends, and a new seed pair is chosen.

This is a (one-step) Markov model; the restriction to single, as opposed to multi-step models can be justified in part by the absence of detectable correlations at two steps, discussed above, and by the reproduction of this absence in the outputs of the one-step model. Different $\mathcal{C}(n,m)$ and combinator choices amount to constraining the $2^{48}\times2^{48}$ transition matrix. The estimation of maximum-likelihood transition probabilities in (hidden) Markov models is often accomplished with a variant of the EM algorithm~\cite{Dempster:1977p16435}; however, even in the simplest model, $\mathcal{C}(1,1)$, the number of parameters ($48^2=2304$) to be estimated is larger than the number of events (1096 fights), and so such iterative methods are unlikely to reliably converge.

On the other hand, determining the parameters directly by searching the full parameter space is impossible. In this exploratory work, we instead make a convenient \emph{Ansatz}. Specifically, we take the elements of $\mathcal{C}(n,m)$ to be equal to the corresponding measurement of the $\Delta P$ between the relevant $n$- and $m$-tuple. In the discussion of results (``How Specific are the Strategies''), we consider a number of alterations from this first guess as a way to assess the flatness of the likelihood and thus to suggest, for future investigations, how to reduce the dimensionality of the parameter space.

Our choice should be reasonably close to the maximum of the likelihood when fights are small and do not grow or shrink too quickly. We find that some choices of strategy class both ``validate'' (approximately reproduce the $\Delta P$ measurements used to specify them) and ``predict'' (reproduce other features of the data that do not directly influence the values of their parameters.)

\begin{figure}[!ht]
\begin{center} \includegraphics[width=4in]{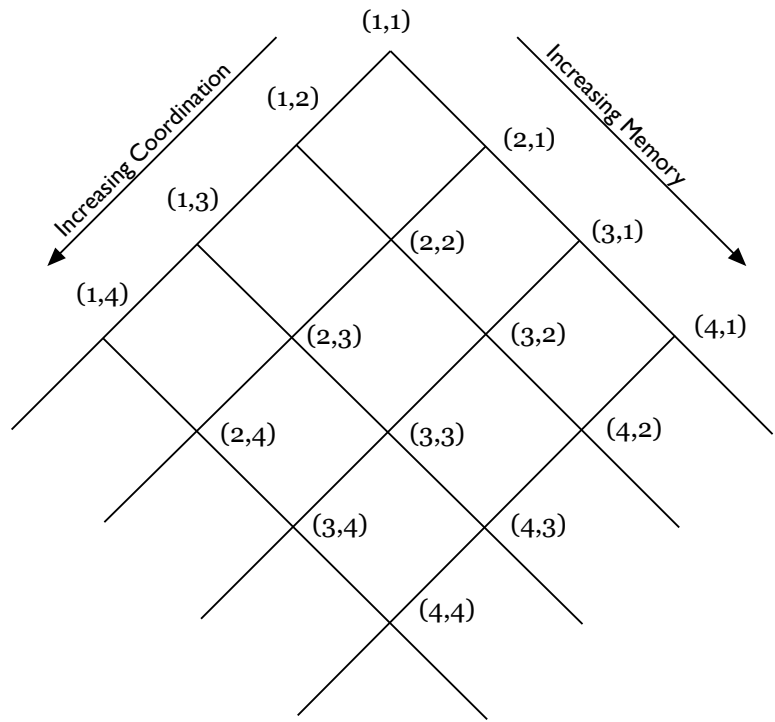} \end{center}
\caption{{\bf Lattice classification of strategy space.} All strategies live in the space of 1-step Markov transition functions. Starting with the simplest model class $\mathcal{C}(1,1)$, we can add individuals to either the first or second fight, systematically building up strategies of increasing complexity based on cognitive, coordination, and computational requirements.}
\label{classification}
\end{figure}

 As shown in Fig.~\ref{classification} we can define a systematic, discrete space of $\mathcal{C}(n,m)$ models that stand in hierarchical relation to one another. Increases in $n$ correspond to an increase in the memory capacity of decision makers. Increases in $m$ correspond to an increase in coordination among individuals. Hence the space defines an hierarchy of memory and information processing requirements. We confine the space of models we consider to those in which $n\leq2$ and $m\leq2$, as the strategies of higher-order models are unlikely to be within the cognitive capabilities of the individuals. In principle, $\mathcal{C}(n,m)$ can be extended systematically to (i) larger values of $n$ and $m$, (ii) include a combinator with more complicated functional dependence, and (iii) accommodate longer timescales, for example, by expanding the dimension of the strategy space from $\mathcal{C}(n,m)$ to $\mathcal{C}(n,m, p)$, and even $\mathcal{C}(n,m, p, z)$, where $p$ and $z$ respectively refer to the second and third time steps from the initial fight.  Later in the Results section, we consider how factors like power \cite{Flack2005b} affect strategy use by individuals. Such factors can be incorporated into the IGT framework but caution is warranted as it is nontrivial to do so systematically; we defer this question to future work.

\section*{Three Hypotheses}
\label{macins}

We consider $\mathcal{C}(1,1)$, $\mathcal{C}(2,1)$ and $\mathcal{C}(1,2)$, with either combinator. Thus each hypothesis has two variants.

$\mathcal{C}(1,1)$ includes only pair-wise decision making strategies that do not require any coordination between conflict participants fighting in the second time-step.  We call this the \emph{Rogue Actor Hypothesis} -- an individual's involvement in a conflict provokes others to become involved in subsequent conflicts. Rejection of this model would suggest that individuals -- by either appearing in many fights themselves or by repeatedly provoking others -- are not the primary cause of fights or cascades. 

$\mathcal{C}(2,1)$ means that an individual decides to participate in a subsequent conflict based on the presence or absence of a particular \emph{pair} of individuals in the previous bout. This model includes triadic-decision making strategies. Rejection of this model would rule out what we call the \emph{Triadic Discrimination Hypothesis} -- individuals make strategic decisions about whether to engage in the present conflict based on who fought with whom previously, and their strategic relation to that pair. 

$\mathcal{C}(1,2)$ means that the decision of a \emph{pair} of individuals to participate in a subsequent conflict is based on the presence or absence of a particular individual in the previous bout. This model includes triadic decision-making strategies that additionally require coordination of participants in the second time-step. Rejection of this model would rule out what we call the \emph{Triadic Coordination Hypothesis} -- individuals jointly decide to fight in a subsequent bout based on the presence of a particular individual in the previous bout, and their strategic relation to that individual. 

Higher-order strategies are in general irreducible -- not decomposable into the products of lower-order strategies. In the language of statistical inference, $\mathcal{C}(1,1)$ is \emph{nested} within the other two strategies; imposing equality constraints allows them to approximate $\mathcal{C}(1,1)$. With these three hypotheses in hand, we can produce simulations of the empirical time series, whose predictions we analyze below. 

\section*{Results} 
\subsection*{Conflict Size}
\label{longfracdiscrim}
We test these hypotheses against each other by simulating conflict dynamics using the $\mathcal{C}(n,m)$ models. We run one simulation for each  $\mathcal{C}(n,m)$+combinator model. We ask how well each of the resulting simulated distributions of fight sizes fits the empirical distribution; the total number of simulated fights is at least 100 times larger than that observed, allowing Monte Carlo estimates of the statistical properties of observable parameters. 

The simulations tell us three things. One is the implication of each $\mathcal{C}(n,m)$ model and its associated strategies for conflict dynamics, including cascade severity. Another is which of the models better reproduces the data, and thus which of the $\mathcal{C}(n,m)$ strategies individuals and subgroups are more likely to be playing in the group. A third insight given by the simulations is how much information the individuals are using, when playing a particular strategy, about other individuals and their interactions. 

\begin{figure}[!ht]
\begin{center} \includegraphics[width=4in]{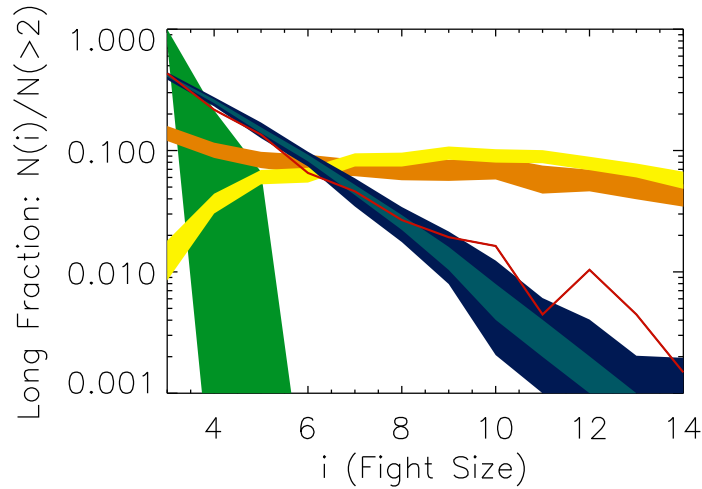} \end{center}
\caption{{\bf Individuals play triadic, not pairwise, strategies, and it is this triadic decision-making that produces turbulent periods.} This plot shows the distribution of fight sizes in the real data (red line) and the simulated distributions under each $\mathcal{C}(2,1)$ hypothesis. We plot the ``long fraction,'' the number of fights of a certain size, divided by the number of fights larger than two participants. In {\bf green} is shown the 95\% confidence contours for $\mathcal{C}(1,1)$+{\tt OR}; the model is unable to generate conflicts of sizes much larger than three. The stricter variant, $\mathcal{C}(1,1)$+{\tt AND}, performs even more poorly. In {\bf orange} is shown $\mathcal{C}(2,1)$+{\tt OR}. Its distribution has a significant fraction of conflicts larger than eight individuals. In {\bf yellow} is $\mathcal{C}(1,2)$+{\tt AND}. Even though this is the ``conflict-averse'' variant of $\mathcal{C}(1,2)$, it produces many large fights over time such that the distribution is ``inverted'' and there are more large fights than small fights. $\mathcal{C}(1,2)$ with the more ``conflict prone'' {\tt OR} combinator produces even larger cascades that grow so quickly good statistics become computationally impossible. Dark {\bf blue} is the 95\% contour and light blue is the 68\% contour for the distribution generated by $\mathcal{C}(2,1)$+{\tt AND}, the only model that can capture important features of the data. This triadic strategy cannot be decomposed into pairwise strategies.}
\label{consequences}
\end{figure}

We operationalize conflict size using a measure we call the ``long fraction'' (Fig.~\ref{consequences}). The long fraction is the number of fights of size $i$, divided by the total number of fights larger than two; formally,
\begin{equation}
\label{lf}
LF(i)=\frac{N(i)}{\sum^{48}_{j=3} N(j)},
\end{equation}
where $N(i)$ is the number of fights of size $i$; the maximum fight size of 48 comes from the total number of socially-mature individuals in the group. The long fraction is a measure of cascade severity, showing how large fights can grow due to the combined strategies of individuals and subgroups. We consider only fights larger than two in size in order to reduce the influence of seed pair composition on the analysis.

As shown in Fig.~\ref{consequences}, the most striking feature of the simulations is the vastly different conflict sizes generated by the different strategies. In the cases considered, these differences allow us to quickly rule out certain simple models. Two of the variants we consider, $\mathcal{C}(1,1)$+{\tt AND} and $\mathcal{C}(1,1)$+{\tt OR},  lead to ``anomalous quiescence'' -- few fights are sufficiently motivating to the group to be consequential. Even if a small conflict manages to double in size, it is rarely able to double again.

We find that in three cases the models lead to ``forest fires'' -- conflict expands in a cascade that engulfs the group, with nearly all individuals participating, and refuses to die down. These are $\mathcal{C}(2,1)$+{\tt OR}, $\mathcal{C}(1,2)$+{\tt OR}, and $\mathcal{C}(1,2)$+{\tt AND}. These strategies do not reproduce the data. Since neither combinator for $\mathcal{C}(1,2)$ works, we rule out the Triadic Coordination Hypothesis.

Only $\mathcal{C}(2,1)$+{\tt AND} reproduces the distribution of fight sizes. This supports the Triadic Discrimination Hypothesis -- individuals decide to fight based on their relation to pairs in previous fights. A small surplus of fights in the data at the very largest fight sizes ($\geq 10$) suggests that strategies of other models might come into play at these extremes. 

This might happen, for example, if during turbulent periods individuals form coalitions in response to the perceived coalitions of others --  $\mathcal{C}(2,2)$. However, the frequency with which this model is used is likely to be low given it requires a level of coordination made difficult by constraints imposed by spatial considerations and limited capacity for communication \cite{Miller2002} among the individuals in individual societies. Model $\mathcal{C}(2,1)$ on the other hand does not require coordination.

The Triadic models, and $\mathcal{C}(2,1)$+{\tt AND} in particular, have (formally) many more parameters than the Rogue Actor Hypotheses. A study of the comparative Akaike Information Criterion (AIC) values, an information theoretic criterion that includes a penalty for model complexity, shows that the improvement in goodness-of-fit is sufficient to compensate; this is discussed in detail in Supporting Information. 

As we noted earlier, the autocorrelation function finds no significant fight size correlations; our model also reproduces this feature. Below we consider a wider range of observables to see how well the Triadic model performs.

\subsection*{Conflict Cost}
\label{conflictcost}

We find in our simulations that the different $\mathcal{C}(n,m)$ strategies have different implications for cascade size -- the pairwise strategies produce small cascades, whereas the triadic strategies produce longer cascades, with the conflict prone variants producing the longest. Here we show, using data taken simultaneously with the time series, that in addition to the assumed costs and benefits to individuals from playing a particular strategy (\emph{e.g.}, that triadic strategies allow individuals to strategically respond to the interactions of others, whereas pair-wise strategies allow no such social ``tuning''), there is a group cost to playing strategies that produce large fights.  (We refer the reader to the Empirical Methods for important operational definitions and statistical methods used in this section.) 

We consider two measures of group cost. These measures capture how likely an individual is to receive aggression given the eruption of a conflict in size class $i$. The first is the (population) mean frequency of contact aggression (\emph{e.g.}, tumbling, wrestling, biting) received by group members during fights in size class, $X_i$. The second is the (population) mean frequency of redirected aggression (\emph{e.g.}, aggression directed by a conflict participant to a third party) received by group members during fight in size class, $Y_i$. The total number of fights in size class $i$ is given by $F_i$. The total number of fights in size class $i$ in which individual $j$ receives contact aggression is $x_{ij}$ and redirected aggression, $y_{ij}$. The aforementioned population-level means are then, $X_i = \frac{1}{48 F_i}\sum_j x_{ij}$ and $Y_i = \frac{1}{48 F_i}\sum_j y_{ij}$. For all large fights (fights size $>4$) $L = \sum_{i=5}^{36} F_i$, and the means are given by $X_L = \frac{1}{48 L} \sum_{j} \sum_{i=5}^{36} x_{ij}$, and $Y_L = \frac{1}{48 L} \sum_{j} \sum_{i=5}^{36} y_{ij}$.  For contact aggression received, the fight sizes are  2, 3, 4, and $>$ 4. For redirected aggression received, the fight sizes are 3, 4, $>$4. By definition, there can be no redirected aggression in fights of size two.

Measuring cost with respect to all individuals in the population rather than conditioning the calculation only on the individuals who fight allows us to capture the consequences to the group of variation in the individual proclivity to fight as well as in strategy variation. All else being equal, the population cost of 10 individuals fighting in a group of 10 is higher than the cost of 10 individuals fighting in a group of 100. Second, by considering redirected aggression, we capture how conflict size affects the likelihood that an individual uninvolved in the dispute will be drawn in. 

\begin{figure}[!ht]
\begin{center} \includegraphics[width=3.5in]{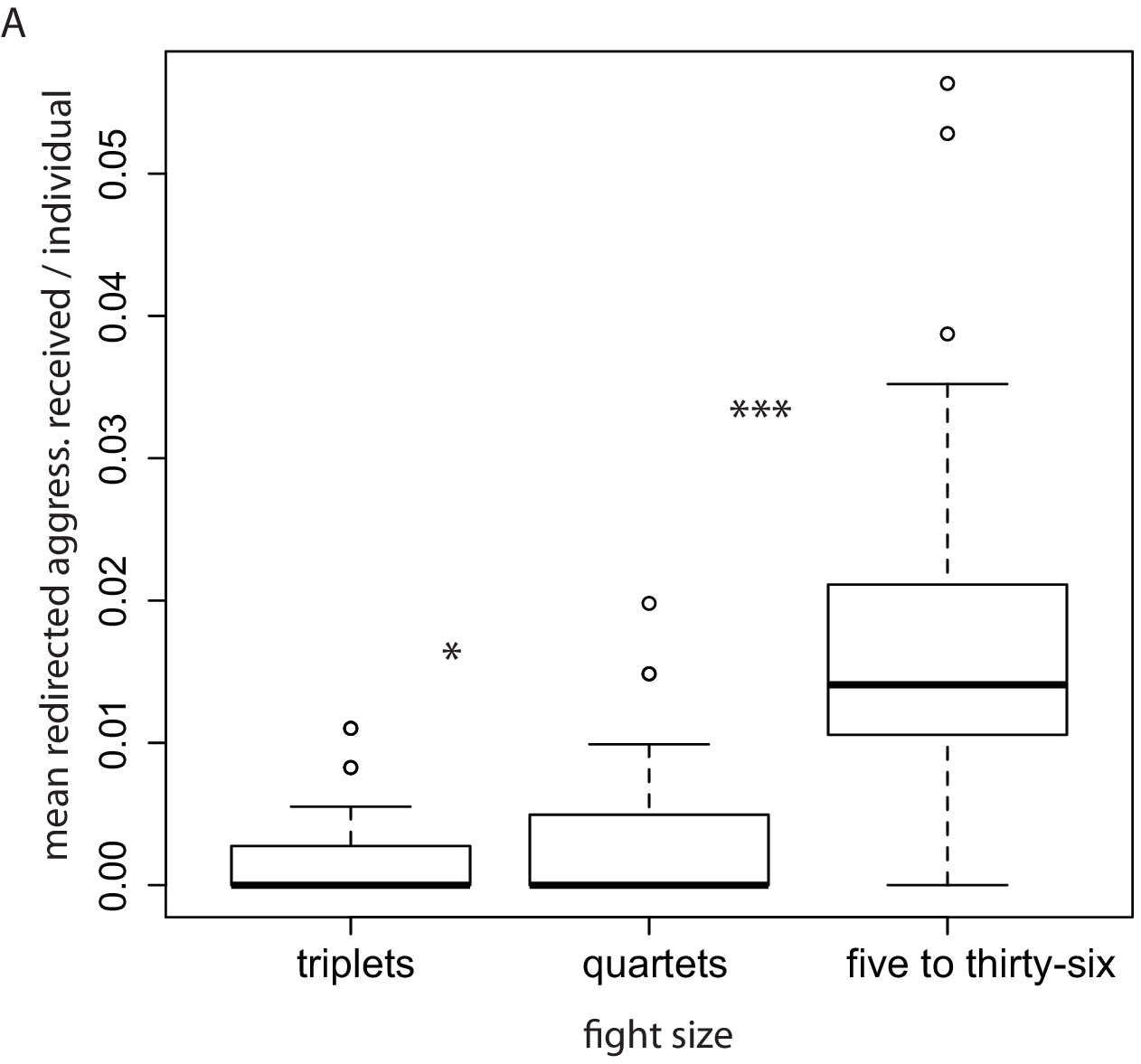} \includegraphics[width=3.5in]{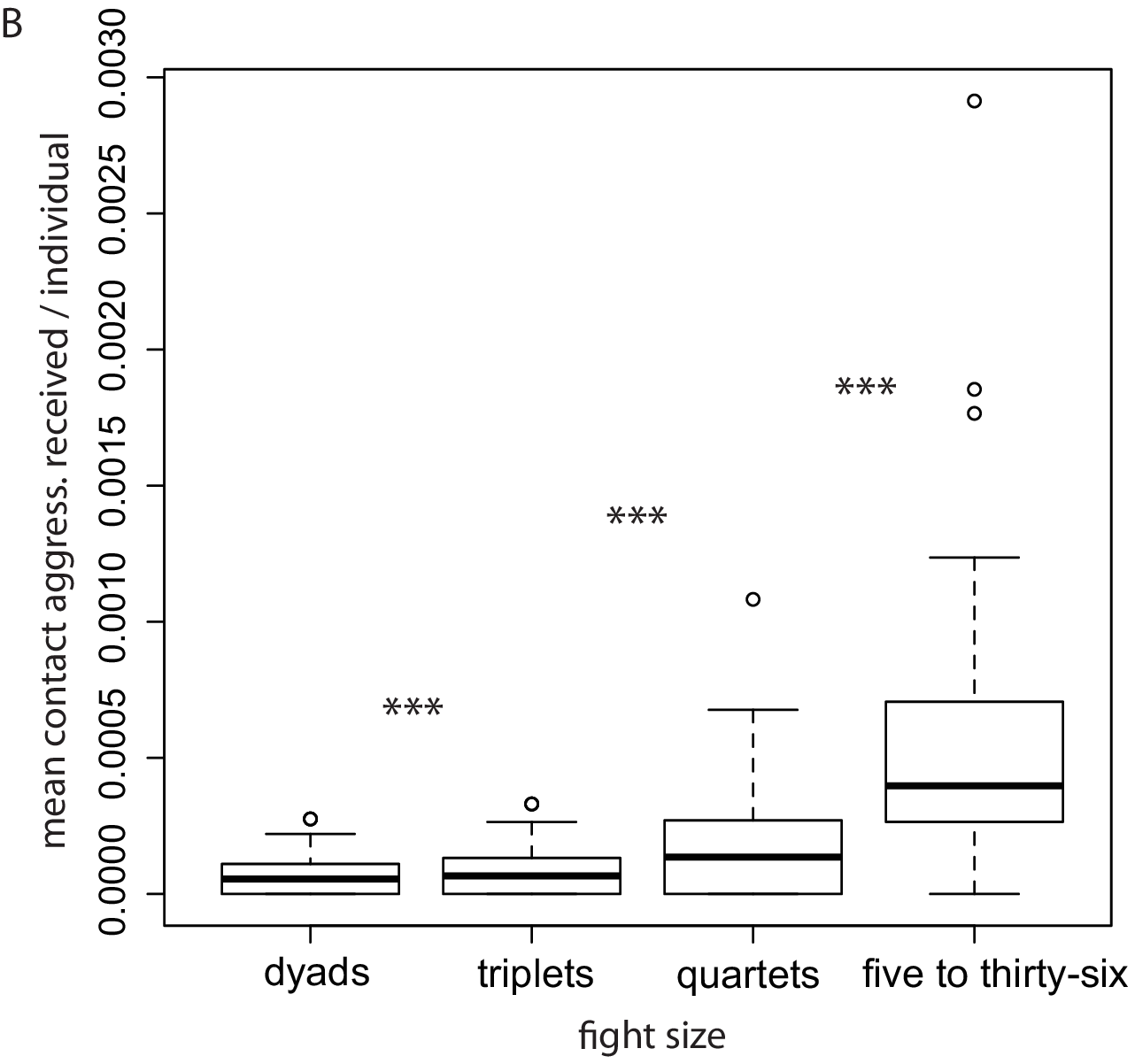}\end{center}
\caption{ {\bf Large fights cost more.} Plot A shows box plots for the mean frequency of redirected aggression received per individual for conflicts of a given size. Plot B shows box plots for the mean frequency of contact aggression received per individual for conflicts of a given size. Conflict sizes were binned so that each category contained an approximately equivalent number of events and to reflect natural categories (\emph{e.g.} pairs and triplets).  The heavy black horizontal line in each plot shows the median ``mean value''. The bottom and top of the box give the 25th and 75th percentiles, respectively. The vertical dashed lines show 1.5 times the interquartile range (roughly two standard deviations). The points are outliers, defined as 1.5 times the interquartile range above the third quartile. Note that redirection, by definition, is not possible in conflicts smaller than triplets.  Adjacent pairs of fight sizes were compared using the Wilcoxon signed ranks test to determine whether the probability of aggression received increases with fight size. The stars indicate the level of significance for differences between adjacent fight sizes.}
\label{cost}
\end{figure}

As shown in  Fig.~\ref{cost} we find, using a paired Wilcoxon signed ranks test, significantly more contact aggression is received by group members when fights are of size 3 than when fights are of size 2 (one-tailed, $n=48, p < 0.001$), when fights are of size 4 than when fights are of size 3 (one-tailed, $n=48, p<0.001$), and when fights are of size $>$ 4 than when they are of size 4 (one-tailed, $n=48, p< 0.001$). Note that the relation between contact aggression received and fight size is nontrivial: aggressors need not use contact aggression and some individuals participate without using or receiving aggression (Methods). Consequently, contact aggression received does not necessarily increase with increasing fight size. Using a paired Wilcoxon Signed Ranks Test we also find significantly more redirected aggression is received by group members when fights are of size 4 than when they are of size 3 (one-tailed, $n=48, p < 0.05$), and when fights are of size $>$ 4 than when they are of size 4 (one-tailed, $n=48, p < 0.001$). 

These results, in conjunction with the results reported in Fig.~\ref{consequences}, suggest that conflict decision-making strategies based on triadic memory are associated with a higher population cost than decision-making strategies based on pair-wise memory. Whether this cost is outweighed by the direct benefits of playing these strategies is a question for future work. 

\subsection*{How Specific are Strategies?}
\label{specific_test}

The model $\mathcal{C}(2,1)$ is triadic; an individual makes a decision to join the present fight depending on the participation of a particular pair of individuals in the previous fight. For each individual, our simulations associate a particular probability with every single pair.

The actual strategies are likely to be far less specific. Cognitive and perceptual constraints mean that a pair might have been perceived as ``Fred and Mary'' or -- at a much lower degree of specificity -- as ``Any Male and Mary.'' A decision-maker's response might also not be so fined graded; instead of a continuum of probabilities, only a finite number of distinct probabilities might be allowed.

In addition to showing the effect of cognitive and biological constraints, studying strategy specificity is important for future work, since by reducing the dimensionality of the space, it could allow direct maximum likelihood searches (see, \emph{e.g.}, \cite{Turner:2008p16417}.)

We consider two variants of $\mathcal{C}(2,1)$+{\tt AND} that are less specific. These are \emph{Shuffled} and \emph{Coarse-Grained}. For clarity, we will sometimes refer to the original model as \emph{Base}.

The \emph{Shuffled} models are alterations of \emph{Base} that re-assign strategies to the group. As with the base model, each individual maintains a static set of strategies from fight to fight. However, the sets used are shuffled compared to the base; we consider three kinds of shuffles. 

\begin{figure}[!ht]
\begin{center} \includegraphics[width=4in]{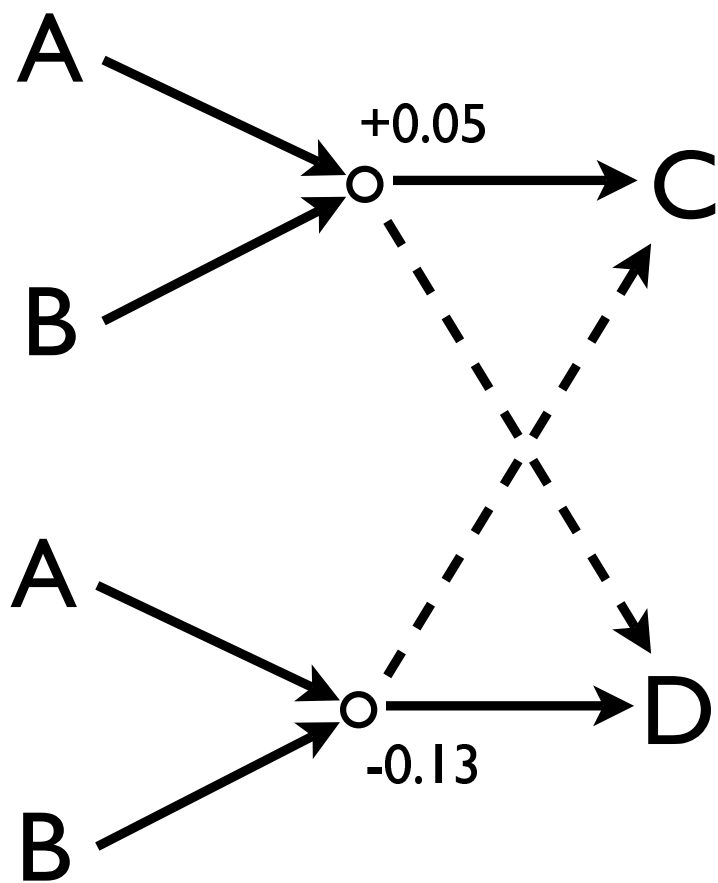} \end{center}
\caption{{\bf A schematic illustration of one of the Triadic tests -- the \emph{Outgoing Shuffle}.} For the incoming pair $AB$, two outgoing names (here, $C$ and $D$) are chosen. The values of the two associated $\Delta P$, $\Delta P(AB\rightarrow C)$ and $\Delta P(AB\rightarrow D)$, are swapped. New names are chosen and the process repeated until all the $\Delta P$s associated with the $AB$ incoming pair have been reassigned. This then is done for all $48\times47/2$ possible incoming pairs, and the resultant $\Delta P$ set used to generate conflict cascades.}
\label{outgoing_shuffle}
\end{figure}

A \emph{Total Shuffle} takes all the combinations $AB\rightarrow C$, and randomly reassigns $\Delta P$ values to them from the original set. An \emph{Outgoing Shuffle} is shown schematically in Fig.~\ref{outgoing_shuffle}. For each of the $48\times47/2$ incoming pairs $AB$, it randomly swaps the $\Delta P$ associated with two outgoing elements. The distribution of the 48 $\Delta P$ values for any particular pair $AB$ remains constant. When possible, the swaps are done between pairs with strategies of opposite sign. An \emph{Incoming Shuffle} is similar, but for incoming pairs; a particular outgoing individual $C$ has the same distribution of $\Delta P$, but they are now randomly associated with different pairs than in the original set.

The \emph{Coarse-Grained} models are, like the base and \emph{Shuffled}, also $\mathcal{C}(2,1)$+{\tt AND}. The particular associations between pairs and individuals are maintained, but the values of $\Delta P$ are now coarse-grained to the nearest of a limited set of $2n+1$ values. For $n$ of one, only three values are allowed: $\Delta P$ equal to the average of all the negative $\Delta P$, equal to zero, or equal to the average of all the (strictly) positive $\Delta P$ values. The example of $n$ of two, with two negative and two positive values of $\Delta P$ allowed, is shown as dotted lines in panel two of Fig.~4 in the Supporting Information. Given the data indicating that Macaque perceptual systems have a logarithmic bias \cite{Nieder2003}, we space the bins logarithmically between the min and max of the positive and negative ranges. 

Testing the coarse-grained models gives a sense of how calibrated an individual's response needs to be to reproduce the data. As $n$ gets larger, the coarse-grained models are closer and closer to the base model in terms of the underlying $\Delta P$ values that dictate the responses of individuals to different pairs. One can consider $n$ a measure of how ``graded'' an individual's responses to a particular pair might be. If $n$ is two, for example, it suggests that individuals class pairs into five categories -- ``don't care'' (zero), ``avoid'' and ``strongly avoid'', and ``join'' and ``strongly join'' -- with no finer distinction.

\begin{figure}[!ht]
\begin{center} \includegraphics[width=4in]{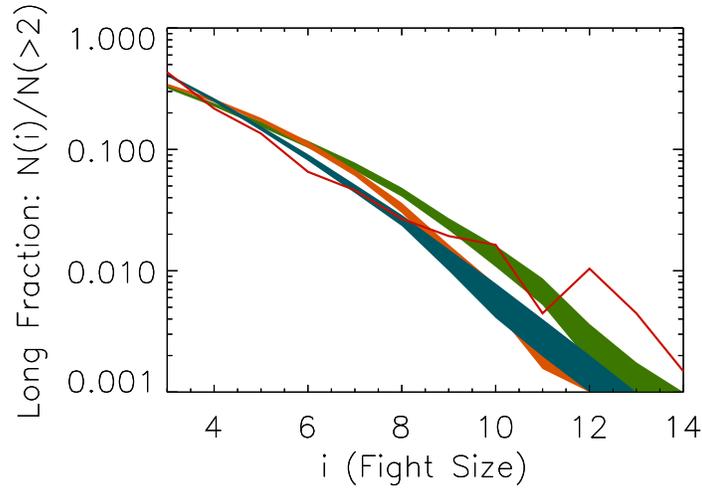} \end{center}
\caption{{\bf The sensitivity of the Long Fraction to $\mathcal{C}(2,1)+${\tt AND} model variants.} 68\% confidence are shown. In blue is the base model. In orange, a simulation based on strategies that have been shuffled relative to the base model (\emph{Total Shuffle}.) In green is a simulation based on strategies where only the incoming pairs have been shuffled relative to the base model (\emph{Incoming Shuffle}.) In red are the data. Both variants of the base, reliant on triadic decision-making, lie nearer to the data than those strategies of Fig.~\ref{consequences}, but still neither are a better fit to the data.}
\label{longfraction_shuffle}
\end{figure}

Earlier in this section, the long fraction alone was sufficient to rule out alternative strategies. The long fractions for the different shuffled strategies, shown in Fig.~\ref{longfraction_shuffle}, also have worse $\chi^2$ values. There are, of course, many more observables than simply the fight size distribution, and we now consider a large set of them. They are (see the Supporting Information) $P(A)$ and $P_c(AB)$, individual and (connected) pair appearance probability; $\bar{n}(A)$ and $\bar{n}(AB)$, average fight size conditional on individual or pair appearance; and $\Delta P(A\rightarrow B)$. In Table~\ref{pearson}, we show the Pearson cross-correlation between the observed data, and the simulations, for the different shuffles and coarse-grainings.

We may also make preliminary estimates of the change in likelihood $\Delta\mathcal{L}$ from the data; we find that the overall likelihood for the parameters drops with either shuffling or coarse-graining. The use of shuffled models also allows us to make a (very preliminary) assessment of the ``true'' number of free parameters in the model, and to penalize the more complicated models; this is discussed in the Model Complexity section of the Supporting Information.

\subsection*{Evidence for Different Strategy Classes}
\label{classes}

The base model for which we find support assumes every individual relies solely on $\mathcal{C}(2,1)$+{\tt AND}. Although it is likely that some of the inconsistency with the data can be removed iteratively through corrections to the $\Delta P$'s as part of a high-dimensional search using an approach similar to Ref.~\cite{Schneidman2006}, it is worthwhile asking whether some subset of individuals and pairs, chosen in a biologically-principled fashion, are better reproduced than others. In other words, are there subsets of individuals that are particularly triadic, and other subsets that either care less about triadic relations or make poorer discriminations? 

We illustrate here how our methods allow one to investigate this question. Individual properties (\emph{e.g.} sex, age, power scores, etc.) can be used to group individuals into categories. We can then ask how well individuals in a particular category are fit by the \emph{Base} model. This can be done by considering for all individuals in the category of interest the two $P(A)$ and $\bar{n}(A)$ measurements, the 94 $P_c(AB)$ and $\bar{n}(AB)$ measurements, and the 95 $\Delta P(A\rightarrow B)$ measurements, and estimating the goodness of fit by computing the associated $\mathcal{L}/n$.

By sorting the individuals into groups based on various extrinsic characteristics, we can determine whether there is evidence for the employment of strategies other than the triadic model of $\mathcal{C}(2,1)$+{\tt AND}. Here, as an illustration of the method, we sort individuals by power score. The power score, discussed in detail in Ref.~\cite{Flack2005b}, is an estimate of how much `consensus' there is among individuals in the group about whether the receiver is capable of using force successfully during fights. Power structure changes the cost of social interactions, facilitating the evolution of intrinsically costly interactions, like policing \cite{Flack2005}, by supporting a  proto-division of labor in which powerful individuals police and low-power individuals do not. Power structure can thus change the strategies individuals play. We expect this variation to influence the extent to which individuals play $\mathcal{C}(2,1)$. 

\begin{figure}[!ht]
\begin{center} \includegraphics[width=4in]{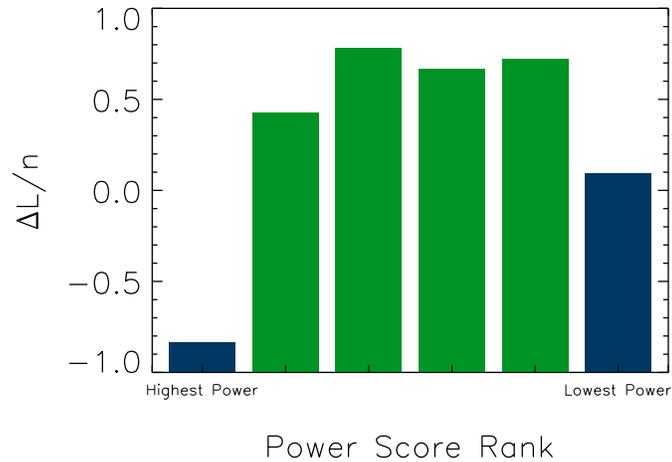} \end{center}
\caption{{\bf Going beyond triadic discrimination.} Overall $\mathcal{L}/n$ as a function of power score, showing how the highest and lowest-power groups are fit least well by the $\mathcal{C}(2,1)$+{\tt AND} strategy assumptions. The 48 individuals are here grouped into units of eight by similarity in power score.}
\label{power_fit}
\end{figure}

We find that the highest power individuals, and the lowest power individuals, are the least-well fit by the data, suggesting that they are using different strategies from those in $\mathcal{C}(2,1)$+{\tt AND} that reproduce much of the behavior of the intermediate-power individuals. This is shown graphically in Fig.~\ref{power_fit}, where the individuals are sorted into groups of eight in order of decreasing power score.

\section*{Discussion}

In this paper we have investigated the causes and properties of conflict in a complex social system. To conduct this investigation we developed a new conceptual and statistical framework applied to conflict time series, which we call \emph{Inductive Game Theory} (IGT.) IGT allows the researcher to computationally extract from data candidate strategies individuals employ to make decisions, and to study using perturbations the effects of alternative strategies on collective dynamics. IGT takes temporally-varying interaction networks as input, and uses these as the basis for a statistical reconstruction of putative, causal networks. These causal network can be used to simulate conditions of conflict, validated against observational data out of sample. Standard, deductive models for the analysis of conflict are not designed to deal with large data sets, and traditionally assume that strategies, payoffs and equilibria can be defined in advance of observation. 

We have applied IGT to a time series in which there are multiple conflicts involving multiple players, and higher-order interactions -- a neglected feature of many gregarious societies, including nonhuman primates, cetaceans, and humans \cite{Harcourt1992}, in which multiple individuals interact at once. We are able to reproduce a number of features of collective behavior, including fight sizes. We discover that the triplet of interacting individuals is an irreducible causal unit for conflict. This is surprising as the pairwise interaction is commonly assumed to be sufficient to explain strategic behavior. 

IGT can be thought of as a complement to a range of statistical, network reconstruction techniques. For example, in genetics, temporal, expression profiles are treated as inputs, and interaction, or transcriptional, networks the desired output~(see, \emph{e.g.}, Ref.~\cite{Bansal2007}.) In IGT we have knowledge of the interactions and seek to derive the collective dynamics, whereas in gene expression, the dynamics are observed, and the interactions are estimated. IGT  is also related also to studies of neural networks that invoke a Ising-model structure to model correlations in the timing of neuronal firings~\cite{Schneidman2006, Tkacik2006}; we differ in that our model invokes a causal process in an out-of-equilibrium system instead of a maximum entropy distribution at fixed temperature. All of these techniques attempt to devise algorithmic approaches to pattern discovery in rich data sets. It is largely the absence of such data in social dynamics that has favored the development of simple models that explain qualitative features of behavior. Whereas IGT has been applied to conflict in this project, there is nothing preventing these ideas being applied to a wider range of collective behavioral sequences, to include prosocial or cooperative behavior, communication, and even coordinated, motor sequences. 

\subsection*{Implications for Social Evolution}
\subsubsection*{Behavior and Cognition}
We find that the primary cause of conflicts in a multiplayer, primate population is individuals responding to the social interactions among others. Neither pair-wise decision-making, nor immediate competition for resources, can account for the conflict patterns we observe in the empirical data. Conflicts are not independent events but are related in time through individual memory for previous conflicts and participants. This effect holds despite peaceful periods, defined by the absence of all overt conflict, separating fights lasting from a few seconds to more than an hour.  We expect memory will play a role in wild populations, although the signal might be noisier as a result of ecological stressors not present in captivity.

Identifying strategies individuals use in deciding to fight requires introducing what we have called a class of minimal models for social reasoning. These models vary in several important respects. One is whether the memory underlying the decision to fight is dyadic or triadic. That individuals primarily use triadic strategies, coupled to the fact that these strategies are not reducible to pair-wise interactions, provides further support for the role of triadic awareness in primate social behavior \cite{Silk1999, Seyfarth1999, Seyfarth2000, Waal2003, Perry2004, Barrett2005}.

A second way in which the models vary, is whether joint action is required. The models we considered were of the form $\mathcal{C}(n,m)$, where $n$ refers to the number and identity of individuals in the previous conflict and $m$ refers to the number and identity of individuals in the conflict. When $m > 1$, $m$ individuals jointly decide to fight in response to $n$. Joint action implies coordination. It is likely that the $\mathcal{C}(n,m)$ models in which $m > 1$ make greater cognitive and spatial demands on the decision-makers than those models for which $m=1$. That we found little support for the $\mathcal{C}(1,2)$ strategy is perhaps explained by these increased cognitive and spatial demands \cite{Miller2002}. 

A third way the models vary is whether the decision-maker is conflict-averse or conflict prone. Our models assume that a decision-maker decides to fight based on its response to individuals or pairs fighting at the previous time step. However, because conflicts can involve multiple pairs, it is possible that the previous fight included both pairs who trigger a ÔjoinÕ response as well as pairs who trigger an ÔavoidÕ response. To deal with these potential decision-making conflicts, we introduced a binary combinator term that specified whether a decision-maker needed a unanimous ``recommendationÕÕ ({\tt AND}) to join or could be pushed over the edge to join by a single recommendation ({\tt OR}). Models with an {\tt AND} combinator we interpreted as conflict-averse strategies whereas models with {\tt OR} combinators we interpreted as conflict-prone strategies. 

Although this binary combinator is crude it roughly captures how a spectrum of temperaments \cite{Clarke2005} and neuro-endocrine profiles might influence decision-making strategies by individuals and their implications for collective conflict dynamics. In the relatively conciliatory \cite{Thierry2000} pigtailed macaque society we study, it is not surprising that the model supported by the data was $\mathcal{C}(2,1)$+{\tt AND} as individuals in this species appear less conflict prone than, for example, individuals in rhesus (\emph{Macaca mulatta}) groups \cite{Thierry2000}. We hypothesize that macroscopic variation in aggression across primate societies \cite{Flack2004} reflects variation in the composition of the fundamental microscopic strategies we have identified inductively. If so, a conflict prone tuning term can explain why in some societies we observe frequent aggression-generated mortality, group fission, and, in captivity, cage wars \cite{McCowan2007}. 

A further cognitive issue raised by these results concerns how much information individuals use to make decisions. Individuals might tune their strategies to individual identity, responding differently to each group member, or more approximately, resolve individuals into classes such as ÔmalesÕ and ÔfemalesÕ. Analogously, behavior can be either discrete or continuous -- with highly tuned responses, or graded responses along the lines of `strongly avoid', `avoid', `join' and `strongly join.' 

The procedure of coarse-graining used in the Results section, "How Specific are the Strategies," suggests that whereas decision-makers have graded rather than continuously varying responses to individuals, they also retain quite fine distinctions between the pairs they react to. These results are consistent with studies of primate cognition showing that individuals can identify other individuals, have the capacity to form numerical representations and discriminate between highly similar vocalizations  \cite{Ghazanfar2005}, can discriminate among emotional states and facial expressions \cite{Parrinpress, Adachi2009}, and have some knowledge of the rank or relative power of other group members \cite{Silk1999, Seyfarth1999, Flack2006b}.

\subsubsection*{Modeling the Role of Conflict in Evolutionary Processes}

There are two primary challenges faced by all complex evolving systems.  One is an uncertain, noisy environment. The other -- the topic of this paper -- is conflict. Conflict arises when the interests of system components -- whether genes, cells, individuals, or states -- are not fully aligned. Conflict is one of the most important social factor shaping the evolution of living systems (for many examples, see Ref.~\cite{Burt2006}) and is thought to have played a prominent role in the evolution of cooperation \cite{Bowles2008, Frank2009}. Some suggest that lack of alignment, or ``frustration'', in many-body systems is the defining feature of \emph{all} complex systems \cite{Sherrington2009}.

Theoretical studies of conflict in particular have proceeded deductively, employing simple models to generate important intuitions about how payoffs select in evolutionary time stable strategies individuals play. In these models there typically is no distinction between evolutionary time and ontogenetic time as the ontogenetic dynamics are either considered transient (timescale too fast to be relevant) or fitness is a simple multiple of payoff. Here we have shown that immediate resource competition does not, at least directly, drive conflict in ontogenetic time in systems with multi-party conflict interactions. Memory for social interactions shapes the strategies individuals employ when deciding to fight, and can generate costly collective conflict dynamics , thereby influencing the evolution of conflict management. The particular strategy used by the individuals in our study group, $\mathcal{C}(2,1)$+{\tt AND}, requires that individuals respond to pairs. Compared to other strategies the individuals could be playing, this triadic strategy induces potentially manageable but not insignificant conflict cascades. We found also found that different strategies have different implications for cascade size and severity, and that larger fights are on average more costly at the population level. These results suggest that the costs and benefits of playing a particular strategy filter back to group members through collective behavior over relatively long timescales of multiple conflicts, as well as directly. It is not clear whether a single integrated payoff can capture these effects.

In addition to the relation between conflict dynamics and resource competition, our work has considered the role of dynamical interaction structure. In evolutionary game theory, interactions are typically pair-wise or, in $n$-person treatments, effectively pair-wise as higher-order strategic interactions tend to be neglected in the mean field \cite{Mesterton2006}. Our finding that the causal unit of conflict dynamics is the triad, not the individual nor the pair, suggests that individual agency has been overemphasized in social evolution. It also suggests that cooperative form and hybrid games \cite{vonneumann1944, Bilbao2000} could come to play a central role when studying competitive and cooperative interactions. A cooperative form game (in contrast to a noncooperative form game -- the standard form in most of evolutionary game theory) is one in which individuals form higher-order units, typically through binding contracts, and play against others through these ``coalitions''. The mathematical definition of coalition is effectively highly correlated constituents; cooperative mechanisms are not required. The interaction structure of these games, as well as that of hybrid games, appears well-suited to studying the stability properties of the strategies our results suggest individuals are playing. 

Finally, using IGT it is possible to computationally extract from data a space of plausible strategies and to study their implications for collective conflict dynamics without positing payoffs. This makes IGT a good complement to standard game theory, which despite its generative power, is well recognized to be weakly tied to natural-system data \cite{Hammerstein2006} and limited by somewhat unrealistic assumptions concerning stationary pay-offs.  

Along with climate change and poverty, conflict is perhaps the most important contemporary challenge to the integrity of human society and to improving individual quality of life. Yet in many respects little is understood about conflict, particularly its causes and dynamics over the life time of an individual. This is because biologists to date have emphasized costs and benefits of conflict in evolutionary time (measured over many generations). The detailed analysis of ontogenetic conflict should provide insights into the behavioral raw material and variability upon which evolutionary dynamics -- both neutral and selective -- operates. 

\section*{Empirical Methods}
\label{ems}

\subsection*{Model System}
Macaque societies are characterized by social learning at the individual level, social structures that arise from nonlinear processes and feedback to influence individual behavior, frequent non-kin interactions and multiplayer conflict interactions, the cost and benefits of which can be quantified at the individual and social network levels \cite {Flack2006, Flack2005, Flack2006b, Flack2005b, Thierry2000, Thierry2004, Flack2004, Johan1993}. These properties coupled to highly resolved data make this system an excellent one for drawing inferences about critical processes in social evolution as well as for developing new modeling approaches that are intended to apply more broadly. 

In this study we focus on one species in the genus, the pigtailed macaque (Macaca nemestrina). The data set, collected by J.C. Flack, is from a large, captive, breeding group of pigtailed macaques that was housed at the Yerkes National Primate Research Center in Lawrenceville, Georgia. Pigtailed macaques have frequent conflict and employ targeted intervention and repair strategies for managing conflict \cite{Flack2005}. The study group had a demographic structure approximating wild populations. Subadult males were regularly removed to mimic emigration occurring in wild populations.  The group contained 84 individuals, including 4 adult males, 25 adult females, and 19 subadults (totaling 48 socially-mature individuals used in the analyses). All individuals, except 8 (4 males, 4 females), were either natal to the group or had been in the group since formation. The group was housed in an indoor-outdoor facility, the outdoor compound of which was 125 x 65 ft. 

Pigtailed macaques are indigenous to south East Asia and live in multi-male, multi-female societies characterized by female matrilines and male group transfer upon onset of puberty \cite{Caldecott1986}. Pigtailed macaques breed all year. Females develop swellings when in \OE strus. 

\subsection*{Data Collection Protocol}
During observations all individuals were confined to the outdoor portion of the compound and were visible to the observer. The $\approx158$ hours of observations occurred for up to eight hours daily between 1,100 and 2,000 hours over a twenty-week period from June until October 1998 and were evenly distributed over the day. Provisioning occurred before observations, and once during observations. The data were collected over a four-month period during which the group was stable (defined as no reversals in status signaling interactions resulting in a change to an individual's power score, see \cite{Flack2006b}). 

Conflict and power (subordination signal) data were collected using an all-occurrence sampling procedure \cite{Altmann1974} in which the compound was repeatedly scanned from left to right for onset of conflict or the occurrence of silent-bared teeth displays (used to measure power, see below). The entire conflict event was then followed, including start time, end time, and the identity of individuals involved as aggressors, recipients, or interveners (see below for operational definitions).  Although conflicts in this study group can involve many individuals, participation is typically serial, making it possible to follow the sequence of interactions. A nearly complete time-series of conflict events is available for each observation period. Breaks in data collection during the day occurred sufficiently rarely (seldom more than once a day), and were sufficiently short (seldom more than fifteen minutes), that results changed little from when correlations were computed assuming no activity during breaks, to not including any fight pairs separated by a break in correlation estimators. We avoided altogether using fight pairs with fights on different days.

Instantaneous scan sampling \cite{Altmann1974} occurred every 15 min for ÔstateÕ behaviours (here, grooming).  

\subsection*{Operational Definitions}
\label{op_def}

Grooming: passing hands or teeth through hair of another individual or plucking the hair with hands or teeth for a minimum of five seconds.

Conflict: includes any interaction in which one individual threatens or aggresses a second individual. A conflict was considered terminated if no aggression or withdrawal responses (fleeing, crouching, screaming, running away, submission signals) occurred for two minutes from the last such event. A conflict can involve multiple pairs if pair-wise conflicts result in aggressive interventions by third parties or redirections by at least one conflict participant. In addition to aggressors, a conflict can include individuals who show no aggression (\emph{e.g.} recipients or third-parties who either only approach the conflict or show affiliative / submissive behavior upon approaching, see \cite{Flack2007}.) Because conflicts involve multiple players two or more individuals can participate in the same conflict but not interact directly. 

Contact aggression: aggression received by one group member from another that involves grappling, tumbling, hitting, slapping, or biting.

Power-disparity: difference between two individuals in their power scores. Power scores for each individual in this study were calculated using a procedure described in \cite{Flack2006b}. In brief, the total frequency of peacefully-emitted subordination signals received by an individual over a given duration (in this case, the study duration, which was approximately four months) is corrected for the uniformity (measured using Shannon entropy) of its distribution of signals received from its population of potential senders (all socially-mature individuals). This equation quantifies how much consensus there is among individuals in the group about whether the receiver is capable of using force successfully during fights. 

Redirected aggression: aggression or threat directed from a conflict participant towards a third-party during or within 5 seconds of the conflict.

Subordination signal: the subordination signal in the pigtailed macaque communication repertoire is the silent bared-teeth display \cite{Flack2007}. Bared-teeth (BT) displays are marked by a retraction of the lips and mouth corners such that the teeth are partially bared. In pigtailed macaques, the SBT occurs in two contexts: peaceful and agonistic SBT see \cite{Flack2007})  Signals in both contexts are highly unidirectional. The agonistic SBT encodes submission. The peaceful variant signals agreement to primitive social contract in which the signaler has the subordinate role \cite{Flack2007}. The network of SBT interactions encodes information about power structure \cite{Flack2006b}.

\subsection*{Statistical Analyses of Empirical Data}
\label{statemp}
In the results of the main paper, we presented results obtained using Wilcoxon Signed Ranks Tests on two measures of cost, contact aggression received and redirected aggression received. We preformed multiple (three for contact aggression received and two for redirected aggression) independent Wilcoxon tests per cost measure instead of one overall Friedman test (nonparametric version of repeated measures) per measure because the \emph{post hoc} planned comparison tests associated with the Friedman test typically do not have enough power to detect differences across treatments. We performed nonparametric tests rather than parametric tests because our data violated the homogeneity of variance assumption. 

\section*{Author Contributions}
Conceived of and designed the simulations: SD, JCF, DCK. Wrote the simulation: SD. Designed the data collection protocol and collected the data: JCF. Analyzed the simulation data SD. Analyzed empirical data JCF. Wrote the paper: SD, JCF, DCK.

\section*{Ethics Statement}

The data collection protocol was approved by the Emory University Institutional Animal Care and Use Committee and all data were collected in accordance with its guidelines for the ethical treatment of nonhuman study subjects.

\section*{Acknowledgments}
SD, JCF and DCK were supported during this project by NIH 1 R24-RR024396-01, NSF NSCC 0904863, and the SFI James S. McDonnell Robustness Program. We thank our three referees for detailed comments on our manuscript. We thank D. Eric Smith (Santa Fe) and Paul Sheridan (Tokyo Institute of Technology) for discussion, and Frans de Waal for support during data collection and the staff of YNPRC for help with data collection.
\newpage


\begin{table}[!ht]
\caption{{\bf Pearson correlation coefficients for the model variants.}}
\begin{tabular}{ l | c | c | c | c | c}
Model & $P(A)$ & $P_c(AB)$ & $\bar{n}(A)$ & $\bar{n}(AB)$ & $\Delta P$ \\ 
$\rho$ & & & & & $A\rightarrow B$ \\ \hline
\emph{Base} & 0.62 & 0.51 & 0.80 & 0.37 & 0.50 \\ \hline
Shuffled & & & &  \\
\emph{Total} & 0.013 & -0.007 & 0.17 & 0.20 & -0.026 \\
\emph{Outgoing} & 0.17 & 0.034 & 0.12 & 0.19 & 0.035 \\
\emph{Incoming} & 0.75 & -0.053 & 0.036 & 0.13 & -0.15 \\ \hline
Coarse Grained & & & & \\
$n=1$ & 0.11 & 0.43 & 0.34 & 0.27 & 0.42 \\
$n=2$ & 0.59 & 0.47 & 0.60 & 0.34 & 0.47 \\
$n=4$ & 0.49 & 0.49 & 0.75 & 0.36 & 0.48 \\
\hline		
\end{tabular}
\begin{flushleft} In nearly all cases, the model outperforms the various ``shuffled'' alternatives, indicating that the triadic nature of the strategies is central to conflict dynamics. The effect of coarse graining the strategies is to reduce correlations; as the number of levels increases and thus finer distinctions are made, the effects disappear. The data suggest that $n=2$ (two positive, and two negative, levels) are sufficient to reproduce much of the group structure. \end{flushleft}
\label{pearson}
\end{table}

\clearpage

\begin{flushleft}
{\Large
\textbf{Supporting Information for \\ Inductive Game Theory and the Dynamics of Animal Conflict}
}
\\
Simon DeDeo$^{1,2}$, 
David C. Krakauer$^{1}$, 
Jessica Flack$^{1,3,\ast}$
\\
\bf{1} Santa Fe Institute, Santa Fe, NM 87501, USA 
\\
\bf{2} Institute for the Physics and Mathematics of the Universe, University of Tokyo, Kashiwa-shi, Chiba 277-8582, Japan 
\\
\bf{3} Santa Fe Institute, Santa Fe, NM 87501, USA \& Yerkes National Primate Research Center, Atlanta, GA 30322, USA\\
$\ast$ E-mail: jflack@santafe.edu
\end{flushleft}

\section*{Aggregate Properties of the Time Series}
\label{aggregate}

To explore how memory might influence collective conflict dynamics, we determine the scales of aggregation in our study group at which significant correlations in conflict properties occur. We test for spurious correlations, which can arise without memory if the same individuals repeatedly engage in conflict simply because they have an aggressive disposition or if particular individuals repeatedly agitate others.

We first  consider the simple case of correlations across fights in the number of individuals involved without regard to identity -- for example, whether large fights follow large fights. We find that, for a wide range of such statistics involving both the median and mean, and different nulls, that there are no detectable correlations of this sort. This section elaborates on that finding.

Consider the statistic ``median over-under correlation,'' an average taken over all fight pairs separated by time $\Delta t$, which can be defined and computed with ensemble averages:
\begin{equation}
\label{ouc}
M(\Delta t) = \langle \Theta(N[t]-\bar{N})\Theta(N[t+\Delta t]-\bar{N}) \rangle,
\end{equation}
where the angle-bracket average is taken over the entire dataset but no pairs with fights on different days are considered. Here $\Theta$ is a step-function: -1 for negative argument and +1 for positive argument. In the case where the fight size of one of the pairs happens to be precisely the median value, \emph{i.e.}, the argument is zero, we take the $\Theta$ function to be $(n-m)/(n+m)$, where $n$ is the number of fights with length strictly above the median value, and $m$ the number of fights strictly below the median value. The time, $t$, and lag, $\Delta t$, are in units of fights within a day; for example, $t$ of five means the fifth fight that day.

The average fight size, $\bar{N}$, can computed within a day or across days.  We find evidence for intra-day variation associated with feeding times and onset of evening hours; however, using a time-dependent $\bar{N}$ complicates analysis without changing any qualitative aspects of the results, and so we do not use this correction. Put simply, then $M(\Delta t)$ estimates how likely a fight of above-median (equivalently, below-median) size is expected to be found after a previous fight also of above-median (below-median) size.

Any correlation statistic such as $M$ will, in general, be non-zero for a particular data set -- even if the data were generated by a completely uncorrelated process. In order to determine the significance level of a detection, we must compare to a null model. Here our null model is a shuffled time-series that leaves the internal properties of the fights unchanged. This null has the advantage of efficiently ensuring that detections will be sensitive to purely time-correlated aspects of the data, and not -- for example -- to the ``zero-lag'' correlations that occur within a single fight or on average. In general, we find our statistics insensitive to whether this shuffling is done within a day -- \emph{i.e.}, keeping same day events on the same day but rearranging their order -- or across all days.

\begin{figure}[!ht]
\begin{center} \includegraphics[width=3.5in]{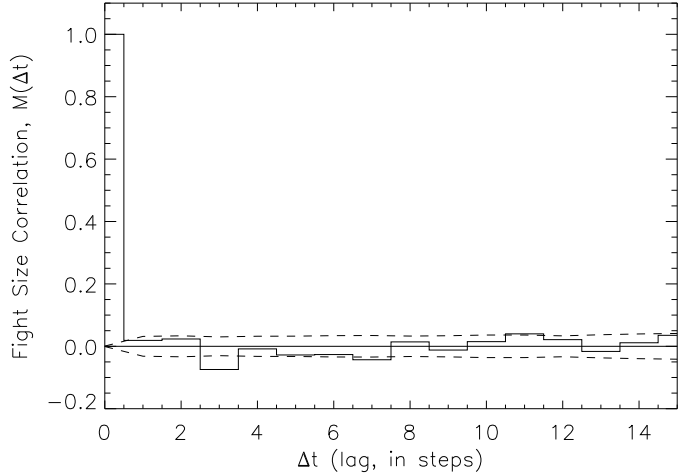} \end{center}
\caption{{\bf Fight size, an aggregate-level variable, does not predict future fight size.} A coarse-grained correlation function, $M(\Delta t)$, of the time series. At lag of zero, $M$ is unity by definition; at non-zero lag, $M$ is indistinguishable from the null model with zero time correlation -- the $68\%$ confidence range for this null is shown as the dashed line. The null model is generated by time shuffling fight bouts.}
\label{coarse}
\end{figure}

As shown in Fig. \ref{coarse}, there is no evidence for memory of previous interactions when considering only fight size data. Conflicts at this coarse-graining appear to occur at random. Furthermore, the correlations of fight sizes, fight durations, and peace durations with themselves and with each other (variations of $M(\Delta t)$ sensitive to, for example, whether a long peace follows a small fight) all appear largely independent of previous events. This result is robust regardless of whether we examine the correlation function in real space or, conversely, in Fourier space where longer baseline trends, such as those involving a small random walk component, might have become apparent. 

The distribution of fight sizes is a central part of our study; as shown in the ``Conflict Cost'' subsection of the Results, it has important consequences for individuals as well as the group as a whole. Larger fights lead to greater amounts of injury and aggression; meanwhile, the consequences of different strategies for fight size distributions are shown in Figs. 4 and 7 of the main paper. Here, we plot the distribution of all fight fight sizes (on a linear scale) in Fig.~\ref{ranked} (overleaf); in Fig.~\ref{ranked_eo} we plot the distribution for the subset of those fights that include the most common fight participant (individual name Eocene, codename ``Eo.'')

\begin{figure}[!ht]
\begin{center} \includegraphics[width=3.5in]{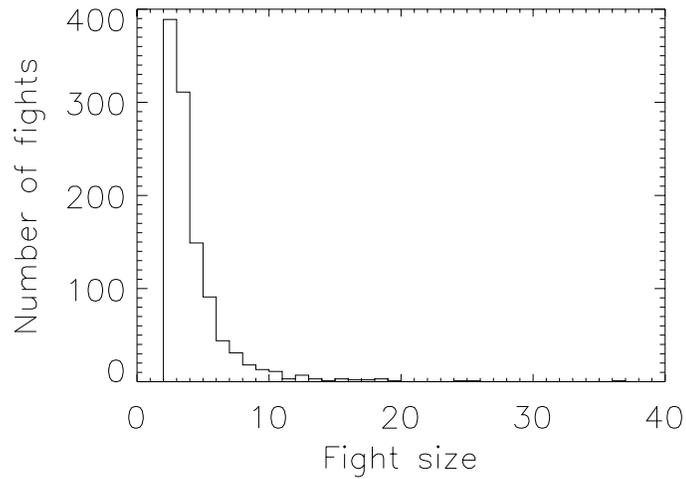} \end{center}
\caption{{\bf Distribution of fight sizes in the data set.} Maximum fight size is 36, mean is 3.7, median is 3.}
\label{ranked}
\end{figure}
\begin{figure}[!ht]
\begin{center} \includegraphics[width=3.5in]{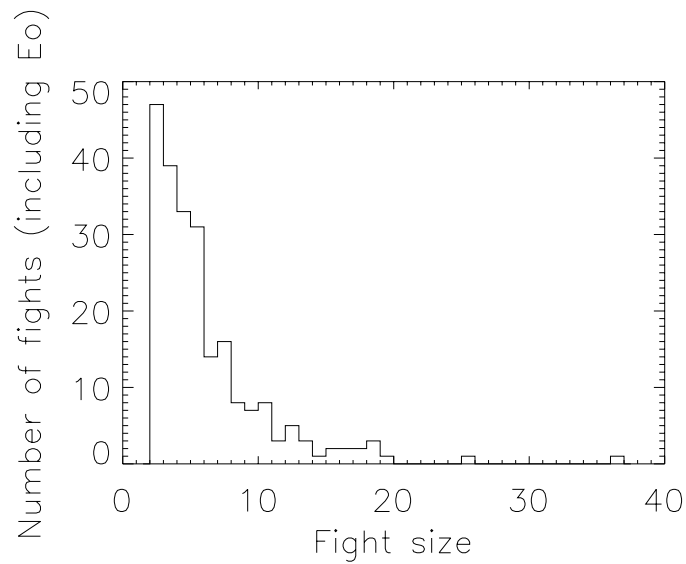} \end{center}
\caption{{\bf Distribution of fight sizes in the data set, for those fights including the most common individual, code name ``Eo'' (Eocene.)} maximum is 36, mean is 5.6, median is 4.}
\label{ranked_eo}
\end{figure}

\section*{Individuals and Subgroups in the Time Series}
\label{individual}

At a finer level of resolution we consider correlations across fights in membership considering all 48 socially-mature individuals. This co-occurrence time-series describes which individuals, pairs, and higher-$n$ groups of individual appeared together in fights, regardless of their behavior.

Because our choice of null conserves the properties of individual fights, $N(A)$ is the same for both the data and the nulls. Given a sufficiently large set of Monte Carlo realizations of the null, we can determine which pairs of individuals have significant time correlations -- \emph{i.e.}, have a $\Delta P$ sufficiently positive or negative that the $\Delta P$ is shared by less than, \emph{e.g.}, 5\% of all Monte Carlo realizations of the null model.

Considering only the group's 48 socially-mature individuals, there are 2,304 possible correlations of the form $\Delta P(A\rightarrow B)$. Results are shown in the (implicitly) causal network in Fig.~2. Many of the strongest correlations are positive, meaning that individuals attract one another to subsequent conflicts, rather than negative, which would mean individuals are inhibited from appearing after certain other individuals appeared.

It is striking the extent to which the absence of signal at the most coarse-grained level of the conflict time-series, Fig.~\ref{coarse}, conceals a rich set of correlations once individual identities are considered, Fig.~2. As mentioned, however, extracting overall significance levels for $\Delta P$ measurements requires caution. For example, since individuals are correlated \emph{within} fights, and these correlations are maintained by the null model, the various $\Delta P$ measurements are not independent of each other.

The effects of these correlations can be seen when comparing the analysis done on the data to that performed on a shuffled version of the data. Taking only those cases where $N(B|A)$ is larger (or smaller) than $95\%$ of $N_\mathrm{null}(B|A)$ found, there are 513 detections of non-zero $\Delta P$ in the data. There are however approximately 480 ``detections'' out of 2304 $\Delta P(A\rightarrow B)$ for the shuffled data -- if the different $N(B|A)$ were perfectly independent, this should be a factor of two lower.

Detections of the positive $\Delta P$ alone are more reliable; we find $296$ $\Delta P$ values larger than zero in the data, compared to an average of $212$ in shuffled datasets, for an overall $p<0.001$. When we come to consider the two-step correlations, the $\Delta P$ can not be distinguished from the noise we expect from the internal correlations of the null model. We defer more detailed discussion of these issues to later work.

\begin{figure}[!ht]
\begin{center} \includegraphics[width=3.5in]{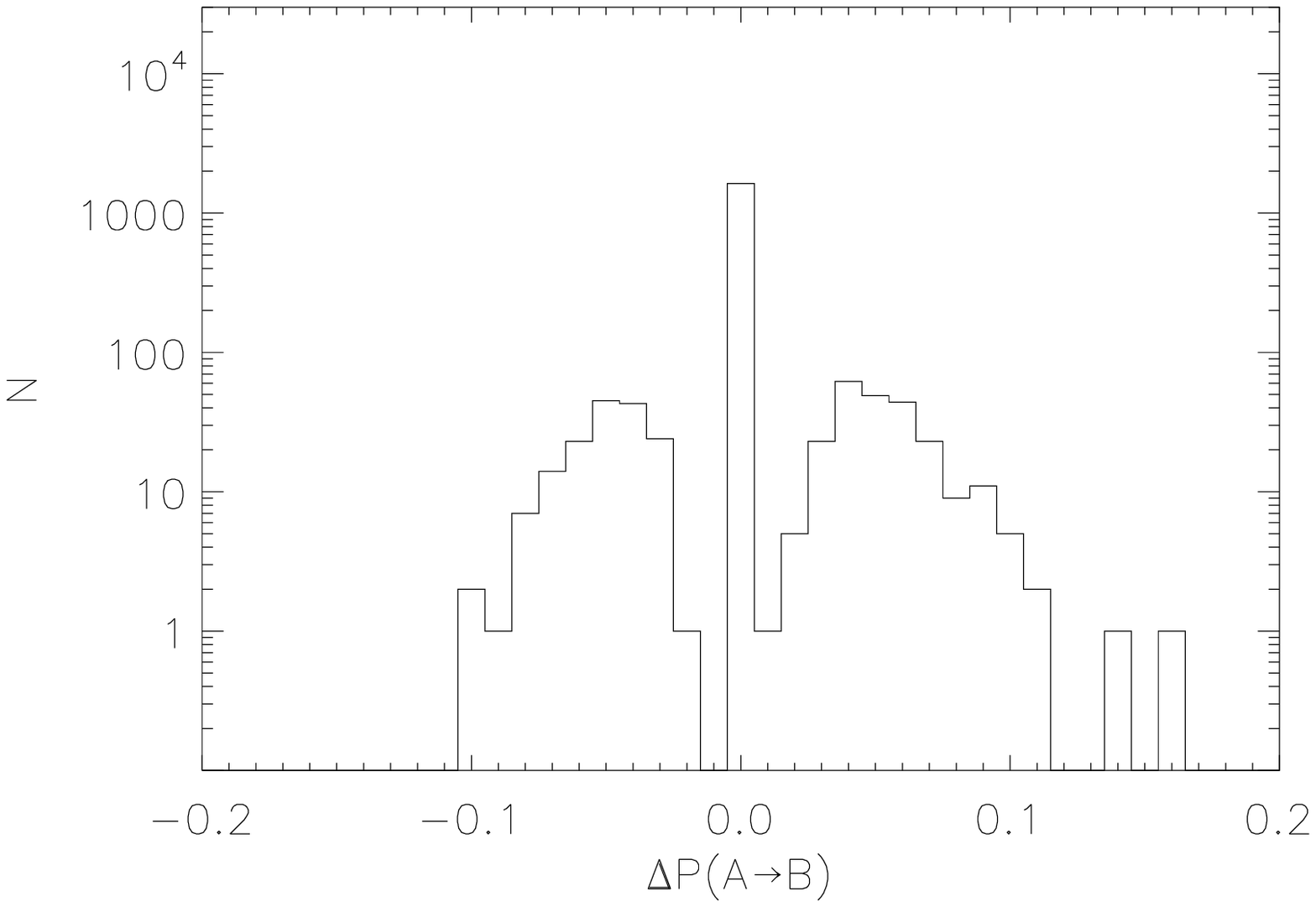} \end{center}
\begin{center} \includegraphics[width=3.5in]{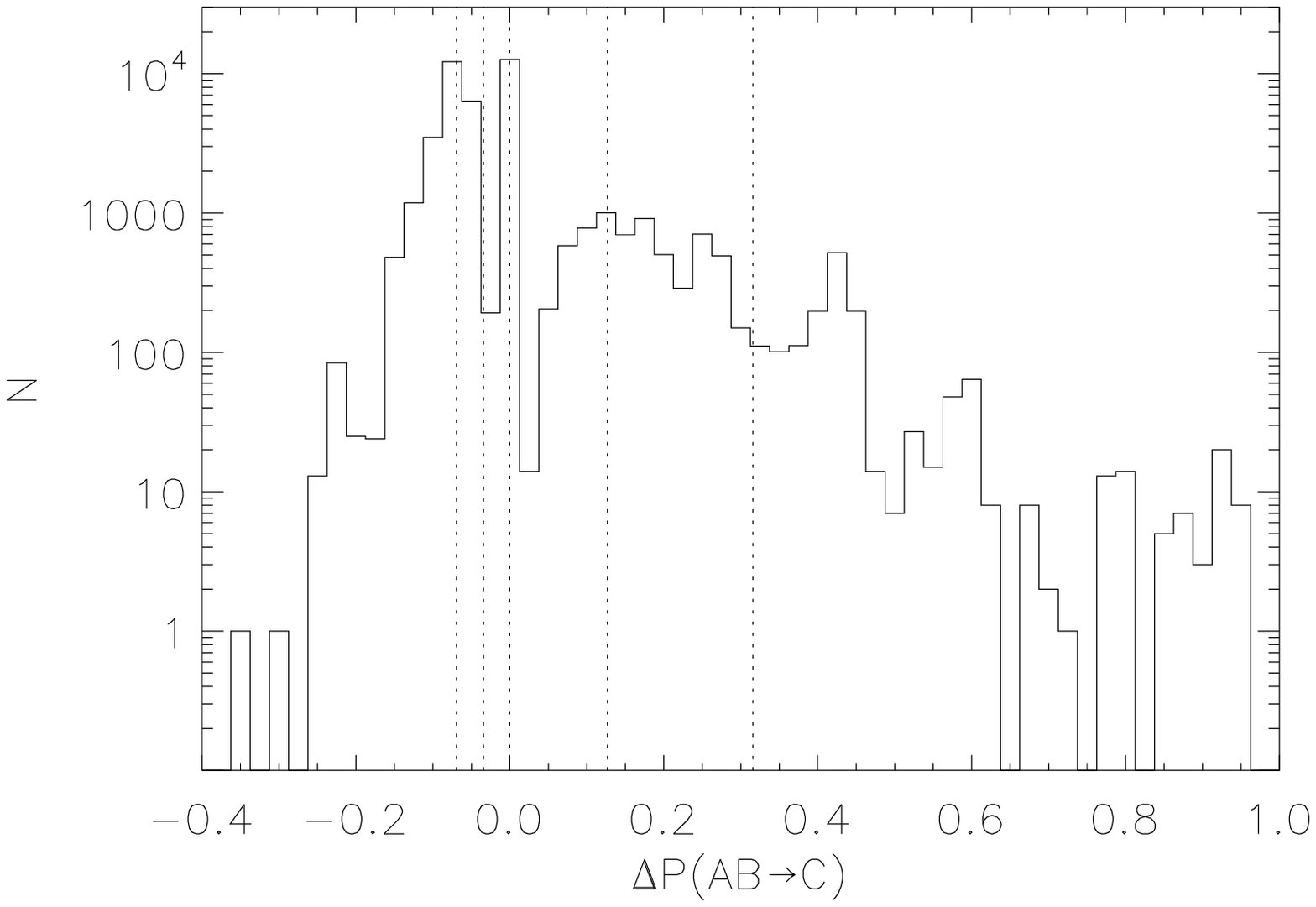} \end{center}
\begin{center} \includegraphics[width=3.5in]{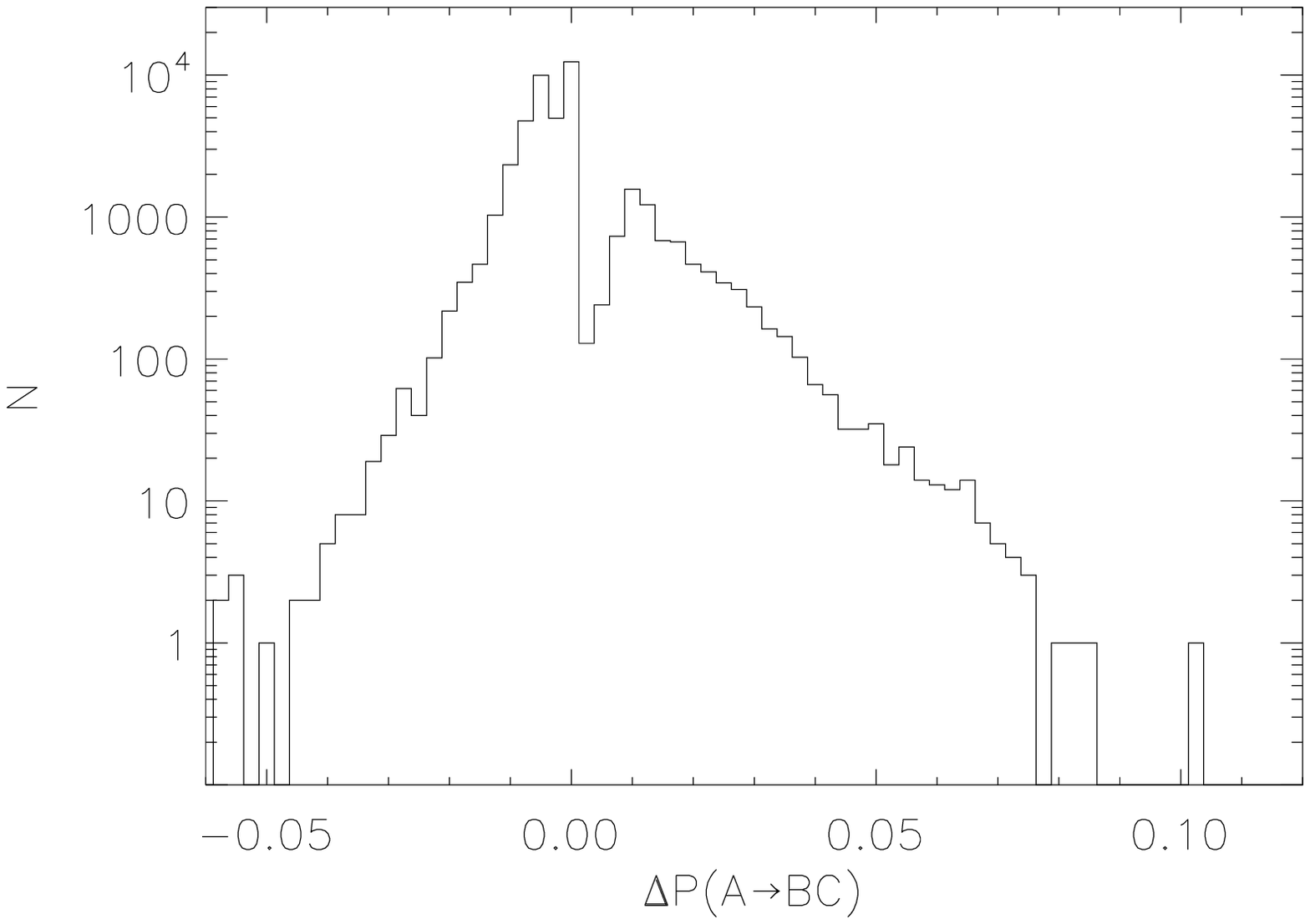} \end{center}
\caption{{\bf The distribution of correlations $\Delta P(A\rightarrow B)$, $\Delta P(AB\rightarrow C)$, and $\Delta P(A\rightarrow BC)$ in the observed data.} These quantify the probability that an appearance of a particular individual or group will be associated with a later appearance (or absence) of another individual or group. The dotted lines, overlaid on the $\Delta P(AB\rightarrow C)$ distribution, are referred to in the strategy specificity subsection of the Results; they indicate the values to which probabilities are coarse-grained at two-level resolution. $\Delta P(AB\rightarrow C)$ values are larger since they are normalized by $N(AB)$ and not $N(A)$.}
\label{distribution_deltap}
\end{figure}
\begin{figure}[!ht]
\begin{center} \includegraphics[width=3.5in]{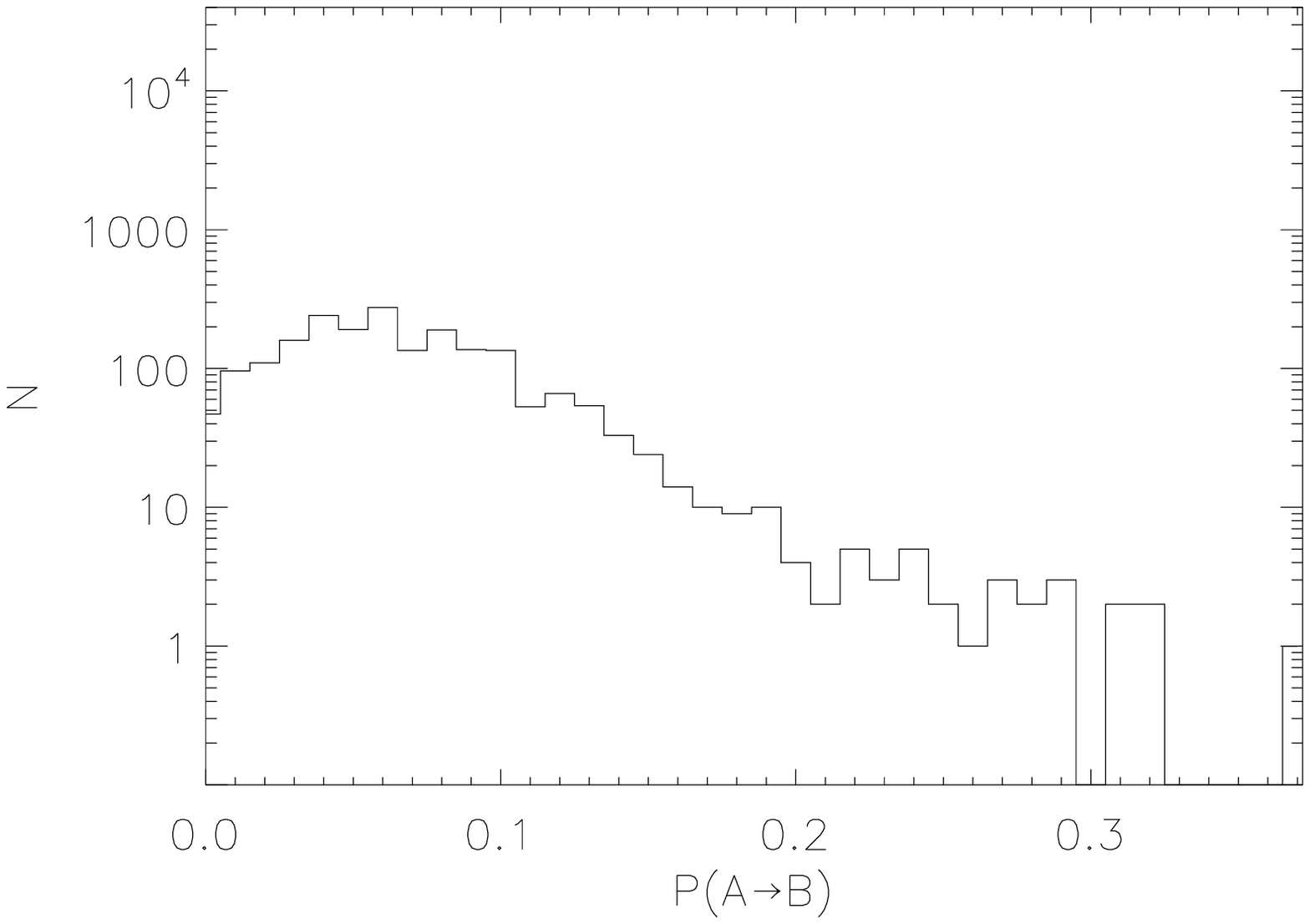} \end{center}
\begin{center} \includegraphics[width=3.5in]{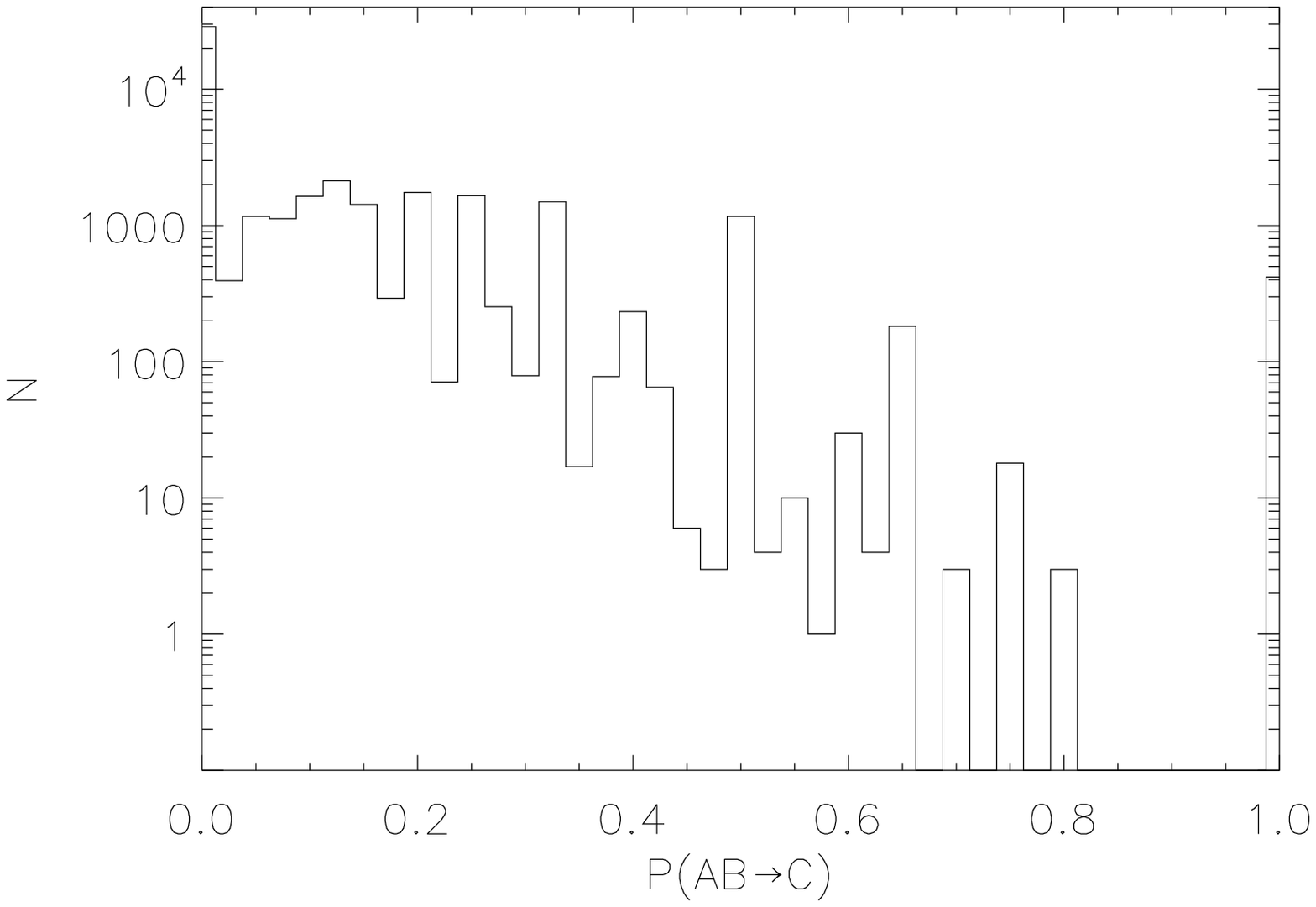} \end{center}
\begin{center} \includegraphics[width=3.5in]{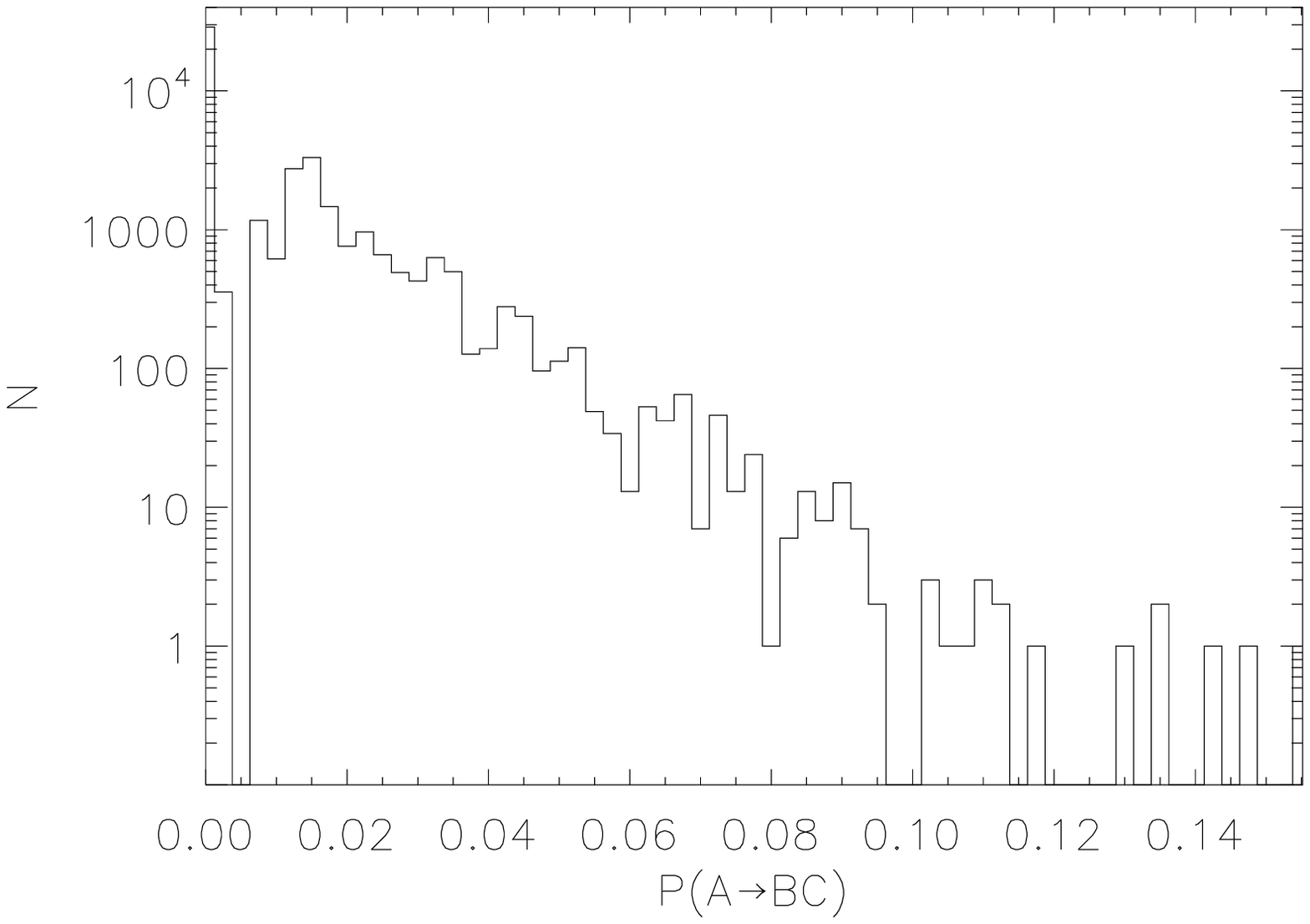} \end{center}
\caption{{\bf As in Fig.~\ref{distribution_deltap}, but now showing the distribution of the absolute probabilities, $P(A\rightarrow B)$, $P(AB\rightarrow C)$ and $P(A\rightarrow BC)$.}}
\label{distribution_deltap_abs}
\end{figure}
The distributions of $\Delta P(A\rightarrow B)$, $\Delta P(AB\rightarrow C)$, and $\Delta P(A\rightarrow BC)$ are shown in  Fig.~\ref{distribution_deltap}; these statistics will play a central role in modeling the group, and in assessing how well the model reproduces the group behavior. In Fig.~\ref{distribution_deltap_abs} we show the distribution of the ``absolute'' probabilities $P$ from which the $\Delta P$ are derived ($P(A\rightarrow B)$ is equal to $N(B|A)/N(A)$, and so forth.)

The null here, recall, is the average from a large Monte Carlo set of null models generated by time-shuffling the series but leaving fight compositions intact. It corresponds to a Markov process where the group has a particular set of conflicts of different compositions that it visits in a biased but uncorrelated fashion. It can be thought of as drawing a fully specified fight from an urn without replacement. 

It is possible to consider relaxing the null model further, so that, for example, individuals join fights with a probability independent of the other members of the group; in this case, one assembles a fight by drawing, with probabilities $P(A)$, $N$ individuals, where $N$ might be, for example, drawn from the observed distribution of fight sizes. Such a model destroys many of the correlations, internal to a fight, that are detectable in the time-averaged properties of the data.

\begin{figure}[!ht]
\begin{center} \includegraphics[width=3.5in]{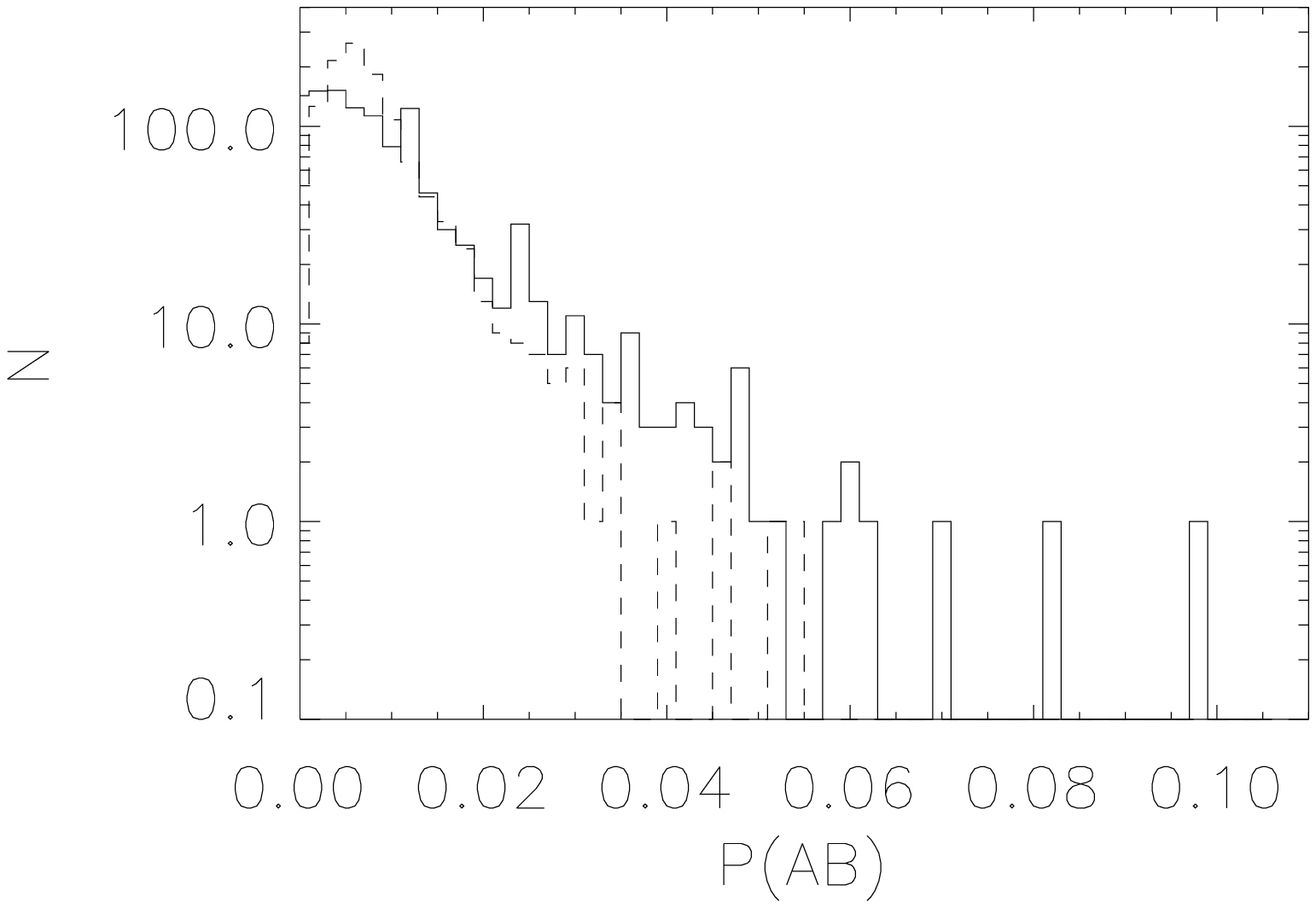} \end{center}
\caption{{\bf The distribution of measured joint appearance probability, $P(AB)$ [solid line], and that expected in the case that individuals appear in fights independently of each other [dotted line].} The two distributions are significantly different under the Kolmogorov-Smirnov test.}
\label{pab_null}
\end{figure}
This can be seen in Fig.~\ref{pab_null}, which compares the observed joint probability, $P(AB)$, and the probability expected if individuals appeared independently of each other, $P(A)P(B)$. A two-sided Kolmogorov-Smirnov test finds the two distributions different with $p\ll 10^{-6}$. Such strong rejections of nulls that ignore correlations between individuals are a general feature of our system, which is strongly correlated both within and between fights.

\section*{Simulation Specification}
\label{simspec}

For a strategy class $\mathcal{C}(n,m)$, we specify two things. First the probabilities, associated with a particular $n$- and $m$-tuple~\footnote{Note that we use ``tuple'' here informally, to mean an unordered set of a particular size.}, that the $n$-tuple recommend the $m$-tuple ``join,'' or ``avoid'' the subsequent fight. In our study here, we do not develop algorithms to search and ``fit out'' the many thousand-dimensional space, but make a good first guess to associate the measurable $\Delta P$ from the simulations with these probabilities. A positive $\Delta P$ for a particular $n$ and $m$-tuple pair is read as a probability to generate a recommendation to join. A negative $\Delta P$ is read as the negative of the probability of a recommendation to avoid. 

A $\Delta P$ of zero (\emph{i.e.}, a $\Delta P$ that could not be detected above a particular confidence level) never generates a recommendation; in many cases, the confidence level requirement zeros out $\Delta P$s that may be formally large -- this is because we exclude measurements of a particular $\Delta P$ that rely on two or fewer instances of fight pairs.

We have checked and find our results insensitive to reasonable changes in the $\Delta P$ cutoff -- either in amplitude or in the number of observations required to establish it. The dynamics of conflict are largely driven by the tuples with easily measured, larger $\Delta P$.

For each possible $m$-tuple in the population, we ``roll the dice'' for all the $n$-tuples in the previous fight. For example, for a fight with four participants, there are four $1$-tuples, six $2$-tuples, four $3$-tuples, and one $4$-tuple. If we are considering a strategy of $\mathcal{C}(2,1)$, we are concerned with the six $2$-tuples, and we will have, for each of them, either a recommendation to join, a recommendation to avoid, or no recommendation at all, for each of the 48 possible outgoing $1$-tuples.

The next step is to combine these. We consider two simple choices of combinator. The recommendations can combined with an {\tt AND} function, requiring a `recommendation to join' from all relevant $n$-tuples at the previous step; here `relevant' means ``with a non-zero $\Delta P$''. Or, the recommendations can be combined with an {\tt OR} function, \emph{i.e.}, needing at least one $n$-tuple in the previous time step to recommend `yes.'

An interpretation of the {\tt AND} rule is conservatism -- that individuals who require many recommendations, or recommendations from all previous conflict participants, to join a subsequent fight are conflict averse. In contrast, an interpretation of {\tt OR} is that individuals who require one or only a few recommendations are pushed over the edge more easily -- they are conflict prone.

\begin{sidewaystable}
\caption{{\bf The implications for social stability of different strategies.}}
\begin{tabular}{ l | c | c | c | c | c }
Strategy & Hypothesis & Combinator & Avg. Max. Size & Avg. Length & Consequences \\ \hline
$\mathcal{C}$(1,1) &``Rogue Actor''  &+{\tt OR} & 2.03 & 1.15 & Anomalous Quiescence \\
 & & +{\tt AND} & 2.02 & 1.10 & Anomalous Quiescence \\ \hline
$\mathcal{C}$(1,2) & ``Triadic Coordination''  &+{\tt OR} & $\sim40$ * & $\gg10^2$ * & Forest Fire \\
 & & +{\tt AND} & 19.9 & 108 & Forest Fire \\ \hline
$\mathcal{C}$(2,1) & ``Triadic Discrimination''  &+{\tt OR} & 5.59 & 3.73 & Forest Fire \\
 & & +{\tt AND} & 2.89 & 2.12 & Manageable \\
\hline		
\end{tabular}
\begin{flushleft} The severity of conflict dynamics, either in terms of average maximum fight size, or average cascade length, varies strongly depending on the model. Column six gives the consequences for the group of a particular model: anomalous quiescence (few large fights), forest fires (long cascades and large fights), or ``manageable'' (no extremely large fights). Memory is pair-wise for $\mathcal{C}(1,1)$ and is triadic for $\mathcal{C}(1,2)$ and $\mathcal{C}(2,1)$. $\mathcal{C}(1,2)$ requires that two individuals jointly engage in conflict at the next time step.  $\mathcal{C}(1,1)$ and $\mathcal{C}(2,1)$ do not require joint action. $\mathcal{C}(2,1)$ requires that individuals discriminate pairs rather than just individuals. As $n$ and $m$ increase, cognitive burden increases because individuals must remember more conflict participants. An asterisk indicates that fights grew so large that reliable statistics were computationally infeasible. \end{flushleft}
\label{stratspace}
\end{sidewaystable}
Given both a model and a combinator, we can now simulate the group's conflict dynamics. We start with a seed pair drawn from the distribution of pairs in the data (we take the distribution of ``fights with only two individuals'', rather than $P(AB)$, but find our results largely insensitive to this choice.) We evaluate the strategies for each individual in the group to determine the composition of the next fight. At some point, a particular fight will, through the propagation effects generated by playing particular strategies, lead to no recommendations for subsequent $m$-tuples to join. At that point, the ``cascade'' is terminated, and a new seed is chosen. As shown in Table~\ref{stratspace}, we consider six different possibilities within the simplest subset of cognitively plausible models.

Our model is not exhaustive, and there are a number of ``limit cases.''  Many of them focus on how to handle joint ``avoid'' and ``join'' recommendations for the same $m$-tuple. In the {\tt OR} combinator case, for instance, does the appearance of a single ``avoid'' recommendation force the tuple to avoid the next fight, or is it overruled by a single ``join'' recommendation? We find that the role of the negative $\Delta P$s is very minor; in particular, it is rare that a recommendation to avoid frustrates the appearance of a tuple that received recommendations to join from other tuples. This seems to indicate that non-appearance in a fight is associated more with disinterest, rather than active decisions based on fear or calculation that lead to successful avoidance.

Another limit case is the situation in which an individual receives a recommendation to join, but no other individual does: the individual ``can find nobody who wants to fight.'' This might depict a behavioral reality, or, conversely, it might indicate the need for new dynamical rules; the single individual may be able to generate conflict in ways not detectable by our segmentation of the data into ``fight'' and ``peace'' bouts. This question is necessarily limited by our coarse-graining of the continuous and complex sequence of events -- individual expressions of aggression over a spatially extended area.

\section*{Composition Statistics}
\label{comp_stats}

We use a set of statistics to quantify how well different variants of $\mathcal{C}(2,1)$ reproduce the data. We consider $P(A)$, the frequency of appearance of individual $A$, and $P_c(AB)$, the connected pair correlation,
\begin{equation}
P_c(AB) = P(AB)-P(A)P(B),
\end{equation}
where $P(AB)$ is the frequency of appearance of the pair $P(AB)$. (The use of only the connected part reduces the covariance of the two statistics.)

We also consider $\bar{n}(A)$ and $\bar{n}(AB)$, the size of the average fight in which one finds individual $A$, or pair $AB$, respectively:
\begin{equation}
\bar{n}(A) = \frac{\sum_{i=2}^{48} iN(A, i)}{N(A)\bar{n}}
\end{equation}
and
\begin{equation}
\bar{n}(AB) = \frac{\sum_{i=2}^{48} iN(AB, i)}{N(AB)\bar{n}}
\end{equation}
where $\bar{n}$ is the average fight size overall, $N(A,i)$ is the number of fights that include $A$ of size $i$, and similarly for $N(AB,i)$. Note that we normalize $\bar{n}(A)$ and $\bar{n}(AB)$ to the average fight size so that, for example, $\bar{n}(Q)$ equal to 2 means that the average fight individual $Q$ appears in is twice the size of an ordinary fight.

Finally, we consider the long fraction, introduced in Eq.~4 of the main paper, and the causal statistic $\Delta P(A\rightarrow B)$.

\section*{Likelihood Estimation for Model Comparison}
\label{stat_meth_appendix}

In the main paper we used the Pearson correlation as a way to determine how well different strategies reproduced the group's behavior. That the base model (and coarse-grained versions with $n\geq2$) generally outperformed the other model variants we took as evidence for the importance of the triadic nature of the strategies.

Here, we take a further step towards quantifying the ways in which our model, and competing variants, reproduce the data. We use the (natural) logarithm of the likelihood ratio per measured parameter, $\Delta\mathcal{L}/n$. This gives an estimate of the relative likelihood of the data being produced by one model versus another. If $\Delta\mathcal{L}/n$ is $\alpha$, then, for an average measurement, the base model is $e^{-\alpha n}$ times more likely than the variant in question.

The calculation of $\Delta\mathcal{L}$ requires an estimate of the shape of the likelihood function, $\mathcal{L}$; this is computationally infeasible to produce exactly. We thus settle for a multivariate Gaussian approximation, and estimate the relative log-likelihoods per degree of freedom in this limit. This gives an estimate of how more strongly preferred the base model is to its variants.

Given the predictions of a particular model we would like to quantify how well they reproduce the data. Doing so requires us to estimate how the observables, which we write schematically as $\{x_i\}$, would change, and how those changes would be correlated with each other, were another set of observations made of the same group under sufficiently similar conditions.

Different simulations imply different noise models, with different correlations. We estimate the intrinsic variation of the data from the simulations themselves, taking many sequences of length $N=1096$ fights, and estimating the statistics -- the various $P(A)$, $P_c(AB)$, $\bar{n}(A)$, $\bar{n}(AB)$ and $\Delta P$  -- from them in the same fashion as in the actual data. We thus reduce the problem to estimating the shape of the likelihood function -- the probability that a measurement drawn from a particular model $H$ will take the values $\{x_i\}$, which we write as $L(\{x_i\}|H)$. When there are only a few $x_i$ to measure, the function can be sampled. Yet for the statistics of interest here, $L$ is a function with potentially thousands of dimensions (for example, the $1128$ pairs, in the case of $P_c(AB)$ and $\bar{n}(AB)$, or the 2304 tuples, in the case of $\Delta P(A\rightarrow B)$.)

The standard solution to this problem is to expand the logarithm of the likelihood to second order about the maximum; this amounts to approximating the distribution of the $n$ statistics as a multivariate Gaussian. For a simultaneous measurement of $n$ statistics -- say, for example, the $1128$ pair probabilities of a $P_c(AB)$ measurement -- the second order expansion is
\begin{eqnarray}
\mathcal{L} & = & -\ln{L(\{x_i\}|H)} \nonumber \\
& = & \frac{n}{2}\ln{2\pi}-\frac{1}{2}\det{\Sigma}+\frac{1}{2}\Delta_i\Sigma_{ij}\Delta_j+\ldots, \label{ll}
\end{eqnarray}
where repeated indices indicate summation, $\Delta_i$ is
\begin{equation}
\Delta_i = (x_i-\bar{x}_i),
\end{equation}
and $\bar{x}$, the value the measurements that maximizes $L$, and $\Sigma$, sometimes called the precision matrix, are a set of parameters we estimate from the simulations. The first two terms, the zeroth order part of the expansion, are what ensure that the likelihood $L$ is normalized properly so that $\int L(\{x_i\}) (\prod dx_i)$ is unity.

We can estimate $\bar{x}$ and $\Sigma$ by the average values, and the inverse of their covariance matrix, of a set of $A$ simulation runs. It is required that $A>n$ for the covariance matrix to be invertible. We can then compare two models -- say, \emph{Base} and \emph{Variant} -- by taking likelihood ratios -- which is more convenient to do in log space:
\begin{equation}
\Delta\mathcal{L} = \mathcal{L}(\{x_i^D\}|\mathrm{Base})- \mathcal{L}_{H1}(\{x_i^D\}|\mathrm{Variant}),
\end{equation}
where $\{x_i^D\}$ are measured from the data. The larger (more positive) $\Delta\mathcal{L}$, above, is, the greater the likelihood of the variant; conversely, negative values of $\Delta\mathcal{L}$ are associated with a higher likelihood attributed to the base model.

\begin{table}[!ht]
\caption{{\bf $\Delta\mathcal{L}/n$, log-likelihood per observable, relative to the \emph{Base} model, for the model variants.}}
\begin{tabular}{ l | c |c  | c | c | c | c}
Model &$P(A)$ & $P_c(AB)$ & $\bar{n}(A)$ & $\bar{n}(AB)$ & $\Delta P$ & Overall \\ 
$\Delta\mathcal{L}/n$ & & & & & $A\rightarrow B$ & \\ \hline
Shuffled & & & & & \\
\emph{Total} & -6.0 & -15 & -8.2 & -5.5 & +0.58 & {\bf -6.7} \\
\emph{Outgoing} & -7.9 & +0.78 & +3.2 & -0.06 & -25 & {\bf -6.7} \\
\emph{Incoming} & -0.61 & +0.58 & -13.3 & -8.0 & +0.37 & {\bf -3.5} \\ \hline
Coarse Grained  & & & & &\\
$n=1$ & -7.7 & -7.3 & -9.7 & -2.2 & -3.1 & {\bf -5.3} \\
$n=2$ & -2.6 & -9.7 & -2.3 & -1.6 & -2.2 & {\bf -4.0} \\
$n=4$ & -3.3 & -7.6 & +2.4 & -0.12 & +0.08 & {\bf -2.0} \\
\hline		
\end{tabular}
\begin{flushleft} For observables of pair properties such as $P_c(AB)$, only the top 100 pairs are considered. In general, individual log-likelihoods are negative, indicating that the base model is preferred over the variants for different subsets of the data; the overall log-likelihoods are all negative. \end{flushleft}
\label{deltal}
\end{table}
Table~\ref{deltal} shows the results, for all 48 individuals (in the case of $P(A)$ and $\bar{n}(A)$), the top 100 tuples (in the case of $P_c(AB)$ and $\bar{n}(AB)$), or the top 100 ``ordered pairs with repetitions allowed'' (in the case of $\Delta P(A\rightarrow B)$.) In general, the entries are negative, indicating that the base model has higher likelihood than the shuffled models. In some cases, $\Delta\mathcal{L}/n$ is positive; indicating that one of the variants has a slightly higher-likelihood for a particular class of measurements. There is, however, no variant that consistently outperforms the base model, and the overall log-likelihoods (computed in the approximation that the different classes of measurements are independent) are all negative.

\section*{Model Complexity}
\label{complexity}

Some of the models that fit the data poorly -- in particular, the two variants of $\mathcal{C}(1,1)$ -- have far fewer parameters than our favored model, $\mathcal{C}(2,1)$+{\tt AND}. There are, formally, 54,144 free parameters in the latter model, while the simpler models have only 2,304. In this section, we investigate, in two different ways, whether the improvements in goodness-of-fit, though large, are justified in an information-theoretic sense. 

Our first intuition is that many of the parameters in $\mathcal{C}(2,1)$+{\tt AND} have little influence on the evolution of the system; for example, setting some of the smaller $\Delta P$ values to zero has little effect on the evolution of the system (see the Supporting Information model specification.) Here, we use the notion of Bayesian complexity to estimate the true number of free parameters; we then use an information theoretic criterion to justify our favorable assessment of $\mathcal{C}(2,1)$.

Bayesian complexity (see Ref.~\cite{Spiegelhalter:2002p17153} and, \emph{e.g.}, Ref.~\cite{Kunz:2006p16542,Liddle:2007p16486}) provides one way to measure the ``effective'' number of free parameters, $k_\mathrm{eff}$. It can be written
\begin{equation}
k_\mathrm{eff} = 2(\ln{L(\hat{\theta})}-\overline{\ln{L(\theta)}}),
\end{equation}
where the overbar denotes average with respect to the posterior PDF and $\hat{\theta}$ are the best estimates of the parameters of the model, $\theta$.

Without a fuller exploration of parameter space, we can not estimate $\overline{\ln{L}}$ to great accuracy. However, if we assume that many of our parameters will be degenerate -- \emph{i.e.}, that our likelihood is reasonably flat -- then the average likelihood can be approximated by the likelihood at a single ``average'' position within the \emph{prior} distribution.

If we take as our prior only the distribution of $\Delta P$ values, then an average point in strategy space is given by the \emph{Total} shuffle, described in Results section of the main paper. We then compute the log-likelihood for the 54,144 values of $N(AB\rightarrow C)$ (directly related to the $\Delta P$ values of the model, these are quicker to compute.) Computing the multivariate Gaussian approximation to the likelihood function for such a large set of parameters is computationally infeasible (it would require over $50,000$ simulation runs), and so we approximate the likelihood as diagonal and Poissonian; because many of the observed frequencies are small (of order one detection in a set of 100 simulation runs of 1000 fights each), we use the functional form of the true Poisson distribution and not standard the Gaussian approximation. We can check to see that the measured covariances are Poissonian (\emph{i.e.}, roughly equal to the measured mean); they are, within the error expected for a measured covariance. 

We find with this method that $k_\mathrm{eff}$ is $\sim2000$. Compared to the formal number of 54,144, this back-of-the-envelope calculation suggests that our intuition -- that many of the parameters of the model are underdetermined and are thus not truly observable -- is correct.

In general, adding more parameters will increase the goodness of fit -- regardless of whether an underlying model requires such complexity; one may, for example, be fitting noise and not signal. There number of information-theoretic statistical tools to penalize model complexity. The \emph{Akaike Information Criterion} (AIC;~\cite{Akaike:1978p16484}), based on the relative Kullback-Leibler information, is one of the best known. Discussed in detail in (for example) Ref.~\cite{Burnham:2004p16485,Liddle:2007p16486}, the AIC value is
\begin{equation}
\label{AIC}
\mathrm{AIC} = -2\ln{L}+2K+\frac{2K(K+1)}{N-K-1}
\end{equation}
where $K$ is the number of parameters in the model and $N$ is the number of observables. A lower AIC value is better, and indicates the preferred model (only relative AIC values are meaningful.)

We can now compare $\mathcal{C}(2,1)$ with  $\mathcal{C}(1,1)$; in particular, we can find an lower bound on the AIC for the $\mathcal{C}(1,1)$ model by assuming its number of effective parameters to be much less than the $2000$ found for $\mathcal{C}(2,1)$.

We find that $\ln{L}/N$ is $-1.1$ for $\mathcal{C}(2,1)$+{\tt AND}, and $-1.9$ for $\mathcal{C}(1,1)$+{\tt OR}; the overall difference in log-likelihood between the two models is then $\sim80,000$. The complexity penalty is $2K+2K(K+1)/(N-K-1)\sim4000$. This means that, despite the complexity of added parameters, $\mathcal{C}(2,1)$+{\tt AND} is significantly preferred over the simpler $\mathcal{C}(1,1)$+{\tt OR}.

\clearpage

\end{document}